\documentclass[conference]{IEEEtran}
\IEEEoverridecommandlockouts
\usepackage{acronym}
\acrodef{3gpp}[3GPP]{3rd Generation Partnership Project}
\acrodef{leo}[LEO]{low Earth orbit}
\acrodef{meo}[MEO]{medium Earth orbit}
\acrodef{geo}[GEO]{geostationary Earth orbit}
\acrodef{ut}[UT]{user terminal}
\acrodef{upa}[UPA]{uniform planar array}
\acrodef{csi}[CSI]{channel state information}
\acrodef{ofdm}[OFDM]{orthogonal frequency division multiplexing}
\acrodef{los}[LOS]{line-of-sight}
\acrodef{toa}[TOA]{time-of-arrival}
\acrodef{pace}[PACE]{positioning-aided channel estimation}
\acrodef{mcrb}[MCRB]{misspecified Cramér-Rao bound}
\acrodef{awgn}[AWGN]{additive white Gaussian noise}
\acrodef{crb}[CRB]{Cramér-Rao bound}
\acrodef{cfo}[CFO]{carrier frequency offset}
\acrodef{peb}[PEB]{position error bound}
\acrodef{crb}[CRB]{Cramér-Rao bound}
\acrodef{lb}[LB]{lower bound}
\acrodef{rmse}[RMSE]{root mean squared error}
\acrodef{fim}[FIM]{Fisher information matrix}
\acrodef{tdd}[TDD]{time division duplex} 
\acrodef{fdd}[FDD]{frequency division duplex} 
\acrodef{wmmse}[WMMSE]{weighted minimal mean squared error} 
\acrodef{qcqp}[QCQP]{quadratically constrained quadratic program}
\acrodef{mrt}[MRT]{maximum ratio transmission}
\acrodef{gnss}[GNSS]{global navigation satellite system}
\acrodef{itu}[ITU]{International Telecommunication Union}
\acrodef{pab}[PAB]{Position-Aided Beamforming}
\acrodef{vdb}[VDB]{Vertically Directed Beamforming}
\acrodef{bs}[BS]{base station}
\acrodef{ntn}[NTN]{non-terrestrial networks}
\acrodef{uav}[UAV]{unmanned aerial vehicle}
\acrodef{tdma}[TDMA]{time-division multiple access}
\acrodef{haps}[HAPS]{high-altitude platform stations}
\acrodef{6g}[6G]{the sixth generation}
\acrodef{bcd}[BCD]{block coordinate descent}
\acrodef{bse}[BSE]{beam squint effect}
\acrodef{cp}[CP]{cyclic prefix}
\acrodef{elaa}[ELAA]{extremely large antenna array}
\acrodef{ff}[FF]{far-field}
\acrodef{las}[L\&S]{localization and sensing}
\acrodef{nf}[NF]{near-field}
\acrodef{ris}[RIS]{reconfigurable intelligent surface}
\acrodef{rtt}[RTT]{round-trip-time}
\acrodef{sns}[SNS]{spatial non-stationarity}
\acrodef{swm}[SWM]{spherical wave model}

\acrodef{siso}[SISO]{single-input-single-output}
\acrodef{mimo}[MIMO]{multi-input-multi-output}
\acrodef{ue}[UE]{user equipment}
\acrodef{dmimo}[D-MIMO]{distributed MIMO}
\acrodef{sp}[SP]{scatter point}
\acrodef{nlos}[NLOS]{non-line-of-sight}
\acrodef{tdoa}[TDOA]{time-difference-of-arrival}
\acrodef{am}[AM]{artificial multipath}
\acrodef{an}[AN]{artificial noise}
\acrodef{psd}[PSD]{power spectral density}
\acrodef{pdf}[PDF]{probability distribution function}
\acrodef{aoa}[AOA]{angle-of-arrival}
\acrodef{aod}[AOD]{angle-of-departure}
\acrodef{moo}[MOO]{multi-objective optimization}
\acrodef{qos}[QoS]{quality of service}
\acrodef{sdp}[SDP]{semi-definite programming}
\acrodef{lmi}[LMI]{linear matrix inequality}
\acrodef{sdr}[SDR]{semi-definite relaxation}
\acrodef{rcs}[RCS]{radar cross section}
\acrodef{isac}[ISAC]{integrated sensing and communication}
\acrodef{pdd}[PDD]{penalty dual decomposition}
\acrodef{bcd}[BCD]{block coordinate descent}
\usepackage{color}

\usepackage{pgfplots}
\usepackage{tikz}
\usetikzlibrary{calc}
\makeatletter
\newcommand{\gettikzxy}[3]{%
  \tikz@scan@one@point\pgfutil@firstofone#1\relax
  \edef#2{\the\pgf@x}%
  \edef#3{\the\pgf@y}%
}
\usetikzlibrary{spy,backgrounds}
\usepackage{mathrsfs}
\usepackage{booktabs} % for better table formatting
\usepackage{amsmath}
\usepackage{graphicx} %use graph format
\usepackage{epstopdf}
\usepackage{amssymb}
\usepackage{amsfonts}
\usepackage{amsthm}
\usepackage{cite}
\usepackage{bm,comment}
\usepackage{algorithm}
\usepackage{algpseudocode}

\usepackage{subeqnarray}
\usepackage{subfigure}
\usepackage{multicol}
\usepackage{multirow}
\usepackage{diagbox}
\usepackage{slashbox}
\usepackage{stfloats}
\usepackage{float}
\usepackage{color} 
\usepackage{cases}
\usepackage{lipsum}

\usepackage[colorlinks=true, citecolor=red, linkcolor=blue, filecolor=blue, urlcolor=blue]{hyperref}
\usepackage{geometry}
\allowdisplaybreaks
\begin{document}
\setlength{\textfloatsep}{4pt}

\bstctlcite{IEEEexample:BSTcontrol}
\title{Positioning-Aided Channel Estimation for Multi-LEO Satellite Cooperative Beamforming}
\author{
Yuchen Zhang, Pinjun Zheng, Jie Ma, Henk Wymeersch, \emph{Fellow, IEEE}, and Tareq Y. Al-Naffouri, \emph{Fellow, IEEE}
\thanks{
This work is supported in part by the King Abdullah University of Science and Technology (KAUST) Office of Sponsored Research (OSR) under Award RFS-CRG12-2024-6478, KAUST Global Fellowship Program under Award No. RFS-2025-6844, and the European Commission through the Horizon Europe/JU SNS project Hexa-X-II under Grant Agreement no. 101095759.

Yuchen Zhang, Jie Ma, and Tareq Y. Al-Naffouri are with the Electrical and Computer Engineering Program, Computer, Electrical and Mathematical Sciences and Engineering (CEMSE), King Abdullah University of Science and Technology (KAUST), Thuwal 23955-6900, Kingdom of Saudi Arabia (e-mail: \{yuchen.zhang; jie.ma; tareq.alnaffouri\}@kaust.edu.sa).

Pinjun Zheng is with the School of Engineering, The University of British Columbia, Kelowna, BC V1V 1V7, Canada (e-mail: pinjun.zheng@ubc.ca). The majority of his contributions to this work were made during his Ph.D. studies at KAUST, Thuwal 23955-6900, Kingdom of Saudi Arabia.

Henk Wymeersch is with the Department of Electrical Engineering, Chalmers University of Technology, 41296 Gothenburg, Sweden (e-mail: henkw@chalmers.se).
}}
\maketitle

\begin{abstract}
We investigate a multi-low Earth orbit (LEO) satellite system that simultaneously provides positioning and communication services to terrestrial user terminals. To address the challenges of accurately acquiring channel state information in LEO satellite systems, we propose a novel two-timescale positioning-aided channel estimation framework, exploiting the distinct variation rates of position-related parameters and channel gains inherent in LEO satellite channels. Using the misspecified Cramér-Rao bound (MCRB) theory, we systematically analyze positioning performance under practical imperfections, such as inter-satellite clock bias and carrier frequency offset. Furthermore, we theoretically demonstrate how position information derived from downlink positioning can enhance uplink channel estimation accuracy, even in the presence of positioning errors, through an MCRB-based analysis. To address the limited link budgets and communication rates of single-satellite communication, we develop a multi-LEO cooperative beamforming strategy for downlink transmission that leverages cluster-wise satellite cooperation while maintaining reduced complexity.
Theoretical analyses and numerical results confirm the effectiveness of the proposed framework in facilitating high-precision downlink positioning under practical imperfections, facilitating uplink channel estimation, and enabling efficient downlink communication.
\end{abstract}
\begin{IEEEkeywords}
LEO satellite positioning and communication, channel estimation, MCRB, cooperative beamforming.
\end{IEEEkeywords}

\IEEEpeerreviewmaketitle
\section{Introduction}
The evolution of mobile communications has been driven by a remarkable vision: enabling seamless, high-speed information access from any point on Earth. While generational advances in cellular networks, from 1G to 5G and beyond, have largely achieved this goal in urban and suburban areas where infrastructure deployment is economically viable, significant challenges remain in rural and remote regions. In these underdeveloped areas, reliable communication and internet access often remain either unavailable or prohibitively expensive, contributing to a growing digital divide\cite{divide2021wcm}. The \ac{ntn}-based communications have emerged as a promising solution to bridge global connectivity gaps, leveraging airborne and spaceborne platforms, such as high-altitude platforms (HAPs) and satellites, as aerial access points\cite{6g2024bits,pan2023tmc,nsr2025,tt2025tvt,yixuan2025taes,ni2025jsac}. Among these platforms, \ac{leo} satellites have garnered particular attention due to several advantages over \ac{meo}, \ac{geo} satellites, and HAPs. These advantages include lower path loss and reduced propagation delays, increased flexibility in constellation design, and comparatively lower deployment and launch costs\cite{wcdmLEO2001wcm,shicong2021leo,jie2024arxiv}.

The advancement of \ac{leo} satellite-based global connectivity presents promising opportunities alongside significant challenges, particularly in channel estimation, a crucial aspect for meeting increasingly demanding high-speed communication requirements\cite{kexin2023twc,malek2024snt}. While receivers can effectively mitigate channel aging by compensating for Doppler shift caused by rapid \ac{leo} satellite movement\cite{wcdmLEO2001wcm,kexin2023twc,semiblind2025cl}, the substantial propagation distance compared to terrestrial networks results in severe signal power attenuation. This attenuation particularly impacts pilot-based channel estimation methods, which struggle with low receive power, making effective \ac{csi} acquisition challenging, especially for power-constrained \acp{ut} such as mobile devices in the uplink. On the other hand, despite the reduced orbital altitude compared to \ac{meo} and \ac{geo} satellites, \ac{leo} communication links still span distances vastly exceeding typical terrestrial communication counterparts. This extended signal travel distance, combined with limited onboard power resources, significantly constrains the link budget of single-\ac{leo} satellite-based communication systems, creating a substantial barrier to achieving the ambitious throughput goals of wideband \ac{ntn} communications\cite{6g2024bits,riccardo2011snt,malek2024snt}.

Although satellite channels pose significant challenges for \ac{csi} acquisition, they exhibit distinct characteristics compared to terrestrial channels, particularly strong \ac{los} dominance and pronounced geometric structures~\cite{riccardo2011snt}. These properties are intrinsically tied to position information~\cite{jie2024arxiv}, suggesting that judicious use of \ac{ut} position knowledge could help alleviate channel estimation challenges in \ac{leo} satellite communications, e.g., by enabling partial channel inference through the reconstruction of geometry-dependent components such as steering vectors. This viewpoint naturally raises two key questions: \emph{(i)} how to acquire accurate UT position information efficiently in \ac{leo} satellite networks, and \emph{(ii)} how to leverage such information to improve \ac{leo} channel estimation and, ultimately, communication performance.
On the other hand, to overcome the limited link budget inherent in single-satellite service, multi-satellite cooperative transmission has emerged as a promising solution under diverse design objectives~\cite{halim2022oj,halim2023oj,kexin2024twc,poor2024tsp,meixia2024twc,daeun2024twc,zack2025dislac,asyn2024twc,asyn2025tvt,moewin2025jsac}. Recent works have, for example, designed cooperative transmission schemes to proactively mitigate asynchronous interference in single-carrier/statistical-\ac{csi} settings~\cite{asyn2024twc}, extended such treatments to \ac{ofdm} systems (often resorting to high-complexity \ac{qcqp}-based formulations)~\cite{asyn2025tvt}, or emphasized joint power allocation and user scheduling with heuristic precoder constructions~\cite{moewin2025jsac}. While these advances are highly relevant, they are not directly aligned with the objective of this paper, which uses cooperative beamforming primarily as a controlled baseline to assess the utility of positioning-aided \ac{csi}. Moreover, despite the growing interest in multi-satellite cooperation, the \emph{computational scalability} of cooperative beamforming remains a key bottleneck, especially for resource-constrained \ac{leo} satellite platforms.

The role of position information in terrestrial networks has been widely recognized for enhancing communication performance\cite{locationaware2014spm,nuria2017cm,fanjiang2023twc,moewin2023jsac,gui2024twc}. For instance, \cite{locationaware2014spm} introduces location awareness in 5G networks, demonstrating its potential for synchronizing coordinated communication schemes. Similarly, \cite{nuria2017cm} showcases how position information facilitates beam alignment in millimeter-wave communication. These approaches typically rely on position information from either \ac{gnss}\cite{locationaware2014spm,nuria2017cm} or radio sensing \cite{fanjiang2023twc,moewin2023jsac,gui2024twc}. However, in \ac{leo} satellite scenarios, obtaining position information of terrestrial \acp{ut} through radar-like sensing at \ac{leo} satellites becomes impractical due to excessive round-trip signal attenuation. While \ac{gnss} appears to be a viable alternative, as demonstrated in \cite{moewin2025jsac}, where \acp{ut}' positions are obtained via \ac{gnss} and fed back to \ac{leo} satellites for downlink communication, {the critical impact of positioning errors and the reliance on overidealistic synchronization assumptions remains under-explored and warrant further investigation in practical implementations.}

Concurrent with advances in \ac{leo} satellite communication, there is growing interest in exploring \ac{leo} satellites as potential supplements or alternatives to \ac{gnss} \cite{zak2024aesm,ferre2022cm,pinjun2025LeoRis,Buehrer2023WCM,Buehrer2025TIT}. This interest arises from the inherent advantages of \ac{leo} satellites over \ac{gnss}, including stronger signal reception, extensive constellations, and broad frequency diversity, characteristics that have also partially driven the success of \ac{leo} satellite communications \cite{ferre2022cm,mugen2024tvt,pinjun2024jsac}. Recent research has begun to investigate the integration of \ac{leo} satellite communication and positioning functionalities. For instance, \cite{you2024twc} proposes a massive \ac{mimo} \ac{leo} satellite system that supports simultaneous communication and positioning in the downlink, designing hybrid beamforming to meet their distinct requirements. However, this work primarily focuses on the tradeoff between positioning and communication performance, overlooking their mutual benefits such as the potential of positioning in facilitating channel estimation, which is a critical challenge in \ac{leo} satellite systems. Moreover, it does not account for the inevitable clock bias and \ac{cfo} among \ac{leo} satellites, factors that may significantly affect positioning performance. 

In this paper, we investigate a cluster of \ac{leo} satellites, each equipped with a \ac{upa}, that simultaneously provide downlink positioning and communication services to multiple single-antenna terrestrial \acp{ut}. We aim to address the key challenges that specifically arise in this setting: characterizing positioning performance under practical synchronization imperfections such as inter-satellite clock bias and \ac{cfo}; leveraging user position information to enhance channel estimation while accounting for inevitable positioning errors; {and mitigating single-satellite link-budget limitations through a carefully designed cooperative multi-satellite beamforming scheme with reduced optimization complexity.}

The main contributions of this paper are as follows. 

\begin{itemize}
\item  Building on the distinct variation rates of position-related parameters and random channel gains in \ac{leo} satellite channels, as detailed in Section \ref{sec-sm}, we propose an innovative two-timescale frame structure. Downlink positioning operates on the longer position-coherent timescale, while uplink channel estimation and downlink communication occur on the shorter channel gain-coherent timescale. Through this framework, each process is scheduled according to its variation rate, maximizing resource utilization while preserving the potential to leverage positioning for efficient channel estimation and communication.

\item To move beyond idealized synchronization assumptions commonly adopted in prior works, we extend the use of the \ac{mcrb} framework to explicitly capture the impact of inter-satellite clock bias and \ac{cfo}. Unlike the classical \ac{crb}, the \ac{mcrb} is designed for estimation under model mismatch~\cite{richmond2017spm}. Through this lens, we rigorously quantify how these synchronization-related impairments distort \ac{toa} and Doppler measurements and, in turn, degrade positioning accuracy in practical \ac{leo} scenarios.  

\item Within the proposed \ac{pace} framework, we show how large-timescale positioning errors propagate into small-timescale channel estimation, creating inherent model mismatch. By applying the \ac{mcrb}, we systematically analyze how uncertainty in the positioning stage translates into performance loss in uplink channel estimation, thereby identifying the robustness margins required for practical position-aided designs.

\item We propose a cooperative beamforming baseline for downlink communication, where each \ac{ut} is simultaneously served by all \ac{leo} satellites in a cluster. We adopt the \ac{wmmse} framework and develop a low-complexity solution that strategically exploits \ac{bcd}, strong duality, and matrix-inversion structures to mitigate the otherwise prohibitive computational burden of multi-satellite cooperation. Importantly, this baseline is designed to be computationally practical and to enable a transparent evaluation of the proposed \ac{pace} framework by comparing the performance achieved under nominal (\ac{pace}-estimated) versus ground-truth channel models.

\end{itemize}

This paper is structured as follows. We begin by presenting the system model in Section \ref{sec-sm}. Section \ref{sec-pace} examines both downlink positioning and uplink channel estimation performance through \ac{mcrb} theory, incorporating the \ac{pace}. A cooperative beamforming approach for downlink communications is detailed in Section \ref{sec-coop}. Numerical results are presented in Section \ref{numer_result}, with conclusions drawn in Section \ref{sec-con}.

\emph{Notations:} Scalars are written in regular lowercase, while vectors and matrices are denoted by bold lowercase and bold uppercase letters, respectively. For a vector $\mathbf{a}$, its 2-norm is expressed as $\left\|\mathbf{a}\right\|$. We use $\mathsf{T}$ and $\mathsf{H}$ as superscripts to indicate transpose and Hermitian transpose operations. For a matrix $\mathbf{A}$, $\text{Tr}(\mathbf{A})$ represents its trace. Complex numbers are handled by $\Re \left\{a\right\}$ and $\Im \left\{a\right\}$ for their real and imaginary components. A block-diagonal matrix constructed from matrices $\mathbf{A}_1,\ldots,\mathbf{A}_N$ is written as $\text{blkdiag}\{\mathbf{A}_1,\ldots,\mathbf{A}_N\}$. The Kronecker product is symbolized by $\otimes$. For statistical distributions, $\mathcal{CN}(\boldsymbol{\mu}, \mathbf{C})$ represents a circularly symmetric complex Gaussian distribution characterized by mean $\boldsymbol{\mu}$ and covariance matrix $\mathbf{C}$.

\section{System Model}\label{sec-sm}

\begin{figure*}[t]
		\centering
		\includegraphics[width=1\linewidth]{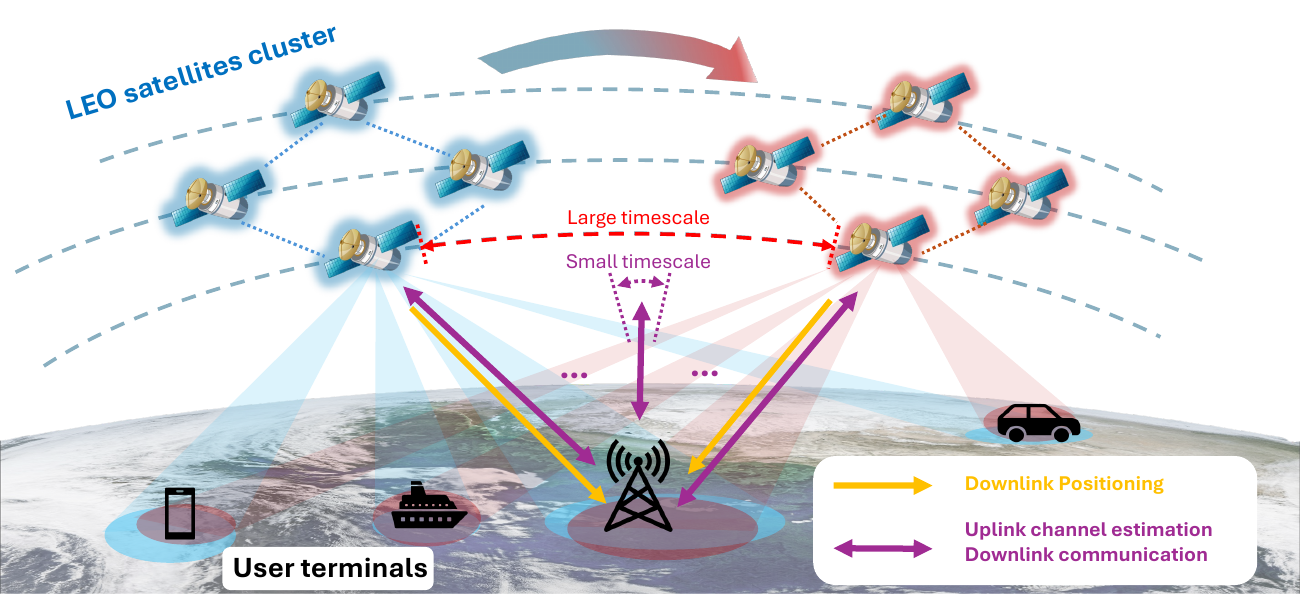}
		\caption{A cluster of LEO satellites traverses orbit while simultaneously delivering positioning and communication services to UTs. Each UT is served by all LEO satellites within the cluster for both functionalities. Positioning operations in the downlink occur over a relatively large timescale, whereas uplink channel estimation and downlink communication are carried out on a smaller timescale. The specific definitions of these timescales will be clarified later.}
		\label{fig_system}
        % \vspace{-5mm}
\end{figure*}

As illustrated in Fig. \ref{fig_system}, we consider a scenario where a cluster of $S$ \ac{leo} satellites serves $U$ \acp{ut}, simultaneously providing downlink positioning and communication services. In this paper, we assume that no handovers occur during operation. The impact of both intra- and inter-cluster handovers is left for future investigation. Each \ac{leo} satellite is equipped with a \ac{upa} comprising $N = N_{\text{h}} \times N_{\text{v}}$ half-wavelength-spaced antennas, where $N_{\text{h}}$ and $N_{\text{v}}$ denote the numbers of antennas in the horizontal and vertical dimensions, respectively. Each \ac{ut} is equipped with a single antenna.\footnote{Note that extending single-antenna \acp{ut} to multi-antenna setting is a promising direction: receiver spatial resolution can help separate signals with distinct Dopplers and delays and mitigate asynchronous interference\cite{asyn2025tvt}. However, a rigorous treatment would also require modifying the positioning model to incorporate array orientation and angle-related measurements, which is beyond the present scope and will be pursued in future work.} While accurate \ac{csi} is essential for achieving efficient downlink communication, channel estimation is challenging in space-borne communication systems due to the limited link budget and short coherence time. These challenges originate from the long travel distance between the satellites and the \acp{ut} as well as the high-speed satellite movement\cite{malek2024snt}. In the following, we first introduce the adopted channel model. Subsequently, we elaborate on a two-timescale property of this model, which can be leveraged to facilitate efficient channel acquisition at \ac{leo} satellites, thereby enhancing downlink communications. Note that handovers in \ac{leo} satellite systems typically occur on the order of tens of seconds to a few minutes, and thus span multiple position-coherent large-timescale frames (see further discussion later). Since the proposed \ac{pace} framework focuses on the integration of positioning and communication \emph{within} each frame, we omit explicit handover modeling to avoid introducing an additional control-plane layer and associated redundancy beyond the scope of this work. This simplifying assumption is also widely adopted in prior studies such as\cite{moewin2025jsac,meixia2024twc,halim2022oj,halim2023oj}.

\subsection{Channel Model}

We consider a satellite communication system operating at relatively high-frequency bands such as the Ku and Ka bands, where the channel is dominated by \ac{los} propagation\cite{riccardo2011snt,poor2024tsp}.\footnote{While the multipath effect can still play a significant role in dense urban environments, \ac{leo} satellite communications are primarily designed to serve suburban, rural, and remote areas, such as oceans and deserts, where terrestrial infrastructure is sparse or nonexistent, and scatterers are greatly reduced. Consequently, \ac{los} propagation becomes even more dominant.} Assume the system employs the \ac{ofdm} scheme with $K$ subcarriers. Let $\Delta f = B/K$ and $T = 1/\Delta f$ denote the subcarrier spacing and symbol duration, respectively, where $B$ is the signal bandwidth. The channel from the $s$-th \ac{leo} satellite to the $u$-th \ac{ut} during the $\ell$-th \ac{ofdm} symbol over the $k$-th subcarrier is expressed as\footnote{Note that, though not further indexed for symbolic simplicity, this expression is viewed within one channel-gain coherent interval (as elaborated later) and $\upsilon_{s,u}$, $\tau_{s,u}$, and $\boldsymbol{\theta}_{s,u}$ are viewed constant for many such intervals.}  
\begin{equation}\label{chan_mod} 
\mathbf{h}_{s,u}\left[\ell,k \right] = \alpha_{s,u} G\left(\theta_{s,u}^{\mathrm{el}}\right)e^{\jmath 2 \pi\left(\ell T \upsilon_{s,u} - k \Delta f \tau_{s,u} \right) }\mathbf{a}\left(\boldsymbol{\theta}_{s,u}\right),    
\end{equation}  
where $\alpha_{s,u}$ is the complex channel gain, $G(\cdot)$ is the symmetric antenna radiation pattern determined solely by the elevation angle\cite{balanis2005antenna}, $\tau_{s,u}$ and $\upsilon_{s,u}$ represent the \ac{toa} and Doppler shift, respectively, and $\mathbf{a}(\boldsymbol{\theta}_{s,u}) \in \mathbb{C}^{N}$ is the steering vector at the \ac{leo} satellite, with $\boldsymbol{\theta}_{s,u} = [\theta_{s,u}^{\mathrm{az}},\theta_{s,u}^{\mathrm{el}}]^{\mathsf{T}}$ being the \ac{aod} composed of both azimuth and elevation angles. Without loss of generality, we assume the \ac{upa} at each satellite is deployed on the XY-plane of its local coordinate system.
%, as illustrated in Fig. \ref{figure_coordinate}. 
Hence, the steering vector can be expressed as  
\begin{equation}  
\mathbf{a}\left(\boldsymbol{\theta}_{s,u}\right) = e^{-\jmath 2 \pi \phi_{s,u}^{\text{h}} \mathbf{n}\left(N_{\text{h}}\right)} \otimes e^{-\jmath 2 \pi \phi_{s,u}^{\text{v}} \mathbf{n}\left(N_{\text{v}}\right)},  
\end{equation}  
where $\mathbf{n}(N) = [0,\ldots,N-1]^{\mathsf{T}}$, $\phi_{s,u}^{\text{h}} = d \cos \theta_{s,u}^{\mathrm{az}} \cos \theta_{s,u}^{\mathrm{el}}/\lambda$, and $\phi_{s,u}^{\text{v}} = d\sin \theta_{s,u}^{\mathrm{az}} \cos \theta_{s,u}^{\mathrm{el}}/\lambda$. Here, $d$ denotes the antenna spacing along each dimension, and $\lambda$ is the wavelength at the central carrier frequency $f_c$.

The complex channel gain can be modeled as  
\begin{equation}\label{chan_gain_formula}
\alpha_{s,u} = e^{\jmath \psi_{s,u}} \beta_{s,u},  
\end{equation}  
where $\psi_{s,u}$ represents the random phase difference between the transmitter and receiver \cite{henk2022cl_i}, and $\beta_{s,u}$ denotes the path loss incorporating both the large-scale path losses and small-scale fading, as described in \cite{3gpp.38.811}. Specifically, the path loss $\beta_{s,u}$, in the dB domain, is characterized as  
\begin{equation}\label{large_loss_component}
-20\log_{10} \beta_{s,u} = \beta^{\text{FS}}_{s,u} + \beta^{\text{SF}}_{s,u} + \beta^{\text{CL}}_{s,u} + \beta^{\text{AB}}_{s,u} + \beta^{\text{SC}}_{s,u}\text{ [dB]},  
\end{equation}  
where $\beta^{\text{FS}}_{s,u}$ denotes the free-space path loss, $\beta^{\text{SF}}_{s,u}$ accounts for shadow fading modeled as a Gaussian random variable, $\beta^{\text{CL}}_{s,u}$ represents the clutter loss, $\beta^{\text{AB}}_{s,u}$ captures atmospheric absorption effects, and $\beta^{\text{SC}}_{s,u}$ accounts for attenuation caused by ionospheric or tropospheric scintillation.  

\subsection{Timescale Analysis and Frame Structure}

\subsubsection{Timescale Analysis} From \eqref{chan_mod}, we observe that the \ac{leo} satellite channel is governed by two categories of parameters:  
\begin{itemize}  
    \item \emph{Large-Timescale Parameters:} These parameters are associated with the positions of the satellites and \acp{ut}, and include the Doppler shift, \ac{toa}, and \ac{aod}. Although \ac{leo} satellites move rapidly, their orbits are \emph{predictable} and well-determined. Thus, the dominant deterministic Doppler component evolves coherently with other geometry-related parameters such as \ac{toa} and \ac{aod}. As long as the \acp{ut}' positions do not change abruptly (which is typically the case in practice), this predictability remains valid over the position-coherent timescale, which is primarily governed by the update rate of the \acp{ut}' positions rather than the satellites' trajectories. Therefore, it is reasonable to assume that large-timescale parameters vary on the order of seconds or longer in typical scenarios.  
    \item \emph{Small-Timescale Parameters:} These parameters consist of random variations in the amplitude and phase of the channel gain. Residual Doppler fluctuations that remain after compensation, due to estimation errors, are also absorbed into this category. Unlike the geometry-determined large-timescale parameters, these small-timescale effects are inherently unpredictable and fluctuate rapidly. Such variations can arise from tropospheric scintillation, which induces sub-second fluctuations in Ka-band \ac{leo} satellite channels depending on atmospheric and orbital conditions \cite{wenwen2010scintillation}. Moreover, even though the \ac{los} component dominates at Ku/Ka bands, nearby mobile scatterers can still generate multipath components that merge with the \ac{los} path, forming a composite channel gain that exhibits small-scale fading with amplitude and phase variations on millisecond timescales \cite{moewin2023jsac,poor2024tsp}.
\end{itemize}

Note that a similar dual-timescale effect has been analyzed and exploited in terrestrial systems \cite{zhao2021twc,fanjiang2023twc}. In \ac{leo} satellite channels, the Doppler shift induced by the high-speed motion of the satellite causes rapid temporal phase rotations. While this effect can be largely mitigated by carrier synchronization at the receiver, small residual errors may remain \cite{wcdmLEO2001wcm}, limiting its impact on communication performance. Nevertheless, we retain the Doppler shift in the channel model since it also provides a valuable measurement for positioning, as will be discussed later. For positioning, the Doppler shift must be explicitly estimated prior to being compensated for communication purposes.

\subsubsection{Two-Timescale Frame Structure}

\begin{figure}[t] 
		\centering
		\includegraphics[width=1\linewidth]{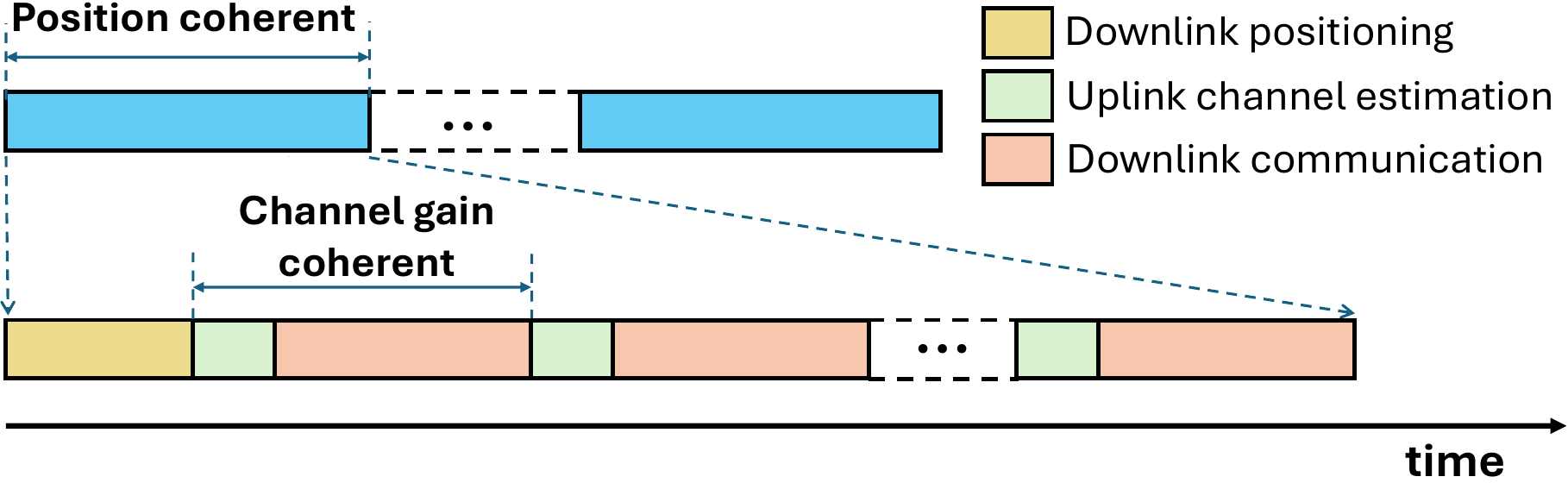}
		\caption{Illustration of the two-timescale frame structure, where each large-timescale frame encompasses multiple small-timescale subframes. The large-timescale frame is governed by the position-coherent time, while the small-timescale subframe is defined by the channel gain-coherent time. }
		\label{fig_frame}
\end{figure} 
Inspired by the timescale analysis, we propose a two-timescale frame structure that exploits the inherent two-timescale property of \ac{leo} channels to enable efficient channel estimation/reconstruction, thus facilitating downlink \ac{leo} transmission. As shown in Fig.~\ref{fig_frame}, the duration of a large-timescale frame is determined by the position-coherent time, which indicates the interval after which the \acp{ut}' position information becomes outdated. Each large-timescale frame begins with an initial subframe for downlink positioning, where \acp{ut}' position estimates can be fed back to \ac{leo} satellites \cite{moewin2025jsac,petar2024ojcomsoc}, followed by multiple small-timescale subframes during which the channel gain remains coherent. We would like to emphasize that the feedback carries only low-rate \ac{ut} position information (a 3D real vector) that remains coherent over the large-timescale interval. This leads to negligible overhead and no stringent latency requirements. Moreover, the dominant uncertainty is the position estimation error (already modeled in our analysis), under which Fig. \ref{up_ce_vs_pos_error} shows that \ac{pace}-based channel reconstruction is robust in practical regimes. Therefore, explicit feedback-link errors are omitted.
Note that \ac{tdd}\footnote{While current NR-\ac{ntn} bands (e.g., n255/n256 \cite{3gpp.38.101-5}) are mainly \ac{fdd}, our \ac{pace} framework relies on uplink–downlink reciprocity and is therefore studied under a \ac{tdd} assumption. Notably, \ac{tdd} for \ac{leo} is explicitly considered by \ac{3gpp} \cite{3gpp.38.811}, has operational precedent in systems such as Iridium Next, and is further examined in recent feasibility studies~\cite{hong2025ttd}. Hence, our \ac{tdd}-based design complements the \ac{fdd} baseline and provides forward-looking insights into reciprocity-enabled schemes for future \ac{ntn} systems.} is adopted to differentiate between uplink and downlink transmissions. The proposed approach leverages uplink-downlink channel reciprocity, which applies uplink channel estimation to optimize downlink communication design. 

% \vspace{-2mm}
To further substantiate the fundamental rationales behind the proposed structure for \ac{leo} downlink communications, we highlight its potential advantages by comparing it with positioning-free uplink channel estimation protocols, where the channel is directly estimated at the \ac{leo} satellites using uplink pilots without requiring position information from \acp{ut}\cite{kexin2023twc,semiblind2025cl}. This approach involves estimating an unknown channel vector/matrix, which can incur significant pilot overhead, especially when large antenna arrays are employed. With the proposed position information-aided framework, vector-type channel estimation can be transformed into scalar-type estimation by exploiting the channel structure. This transformation offers the potential for higher channel estimation accuracy with the same pilot overhead. Alternatively, to achieve equivalent channel estimation accuracy, reduced pilot overhead is required, thus preserving more resources for data transmission.

In the following, we will first conduct a performance analysis based on the two-timescale \ac{pace} and then propose a cooperative beamforming method across multiple \ac{leo} satellites for downlink communication.

\section{Two-timescale Positioning-Aided Channel Estimation}\label{sec-pace}
To characterize its performance limits, we conduct a two-phase theoretical performance analysis\footnote{For both downlink positioning and uplink channel estimation, we theoretically characterize the performance under model mismatch using the \ac{mcrb}, with the aim of revealing the potential of the proposed \ac{pace} framework in enhancing channel estimation. The design of specific estimators is beyond the scope of this paper and is left for future work.} for \ac{pace}. In the first phase, we analyze the downlink multi-\ac{leo} satellite-based positioning performance, while in the second phase, we assess the uplink channel estimation performance. 

Note that model mismatches can arise from various hardware imperfections.\footnote{While impairments such as phase noise and power amplifier nonlinearities are also relevant at Ku/Ka band\cite{kefeng2019tcom}, we focus on clock bias and \ac{cfo} because they directly perturb delay and Doppler, introducing \emph{deterministic biases} across satellites that dominate positioning errors. By contrast, phase noise and nonlinearities mainly act as \emph{stochastic distortions} that reduce effective received power and can, in principle, be mitigated on a single-satellite basis.} First, clock biases are inevitable between satellites and \acp{ut} as well as among satellites themselves. Although sub-nanosecond synchronization can be achieved by equipping \ac{leo} satellites with \ac{gnss} receivers~\cite{kunzi2022precise}, the received signals remain subject to ionospheric and tropospheric delays during propagation~\cite{prol2022position}. These delays, shaped by temperature, pressure, humidity, and the satellite's elevation angle relative to the \ac{ut}, induce unique offsets for each satellite–\ac{ut} pair and effectively translate into \emph{equivalent} clock biases in the signal model. Second, low-cost oscillators and inherent hardware impairments can introduce \ac{cfo}~\cite{Peng2024tvt}.
Many localization algorithms simplify the underlying model by neglecting certain factors to reduce analytical complexity. To capture the impact of this simplification, in the positioning phase we employ a simplified model to estimate unknowns that are actually generated by a more realistic one, which inherently introduces model mismatches. Similarly, in the second phase, position information-aided channel estimation, estimation errors in \ac{ut} positions also lead to mismatches. Consequently, the performance bounds under such conditions can be effectively characterized using the \ac{mcrb}~\cite{richmond2017spm,zack2024globecom}.

\subsection{Large-Timescale Downlink Positioning}
\subsubsection{Signal Model} 
Let $\mathbf{t}_s[\ell,k] \in \mathbb{C}^{U}$ be the pilot (with unit-modulus elements) sent by the $s$-th \ac{leo} satellite over the $\ell$-th symbol and the $k$-th subcarrier. We assume the satellites are multiplexed in the positioning phase through \ac{tdma}\cite{pinjun2024jsac}. The received baseband signal at the $u$-th \ac{ut}, from the $s$-th satellite can be expressed as
\begin{align}
    y_{s,u}\left[\ell,k \right] = &\mathbf{h}_{s,u}^{\mathsf{T}}\left[\ell,k \right] \mathbf{F}_s\mathbf{t}_s\left[\ell,k \right] + z_u\left[\ell,k \right] \notag \\    =&\vartheta_{s,u} \gamma\left[\ell,k\right]\mathbf{a}^\mathsf{T}\left(\bm{\theta}_{s,u}\right)\mathbf{f}_s\left[\ell,k\right]+z_u\left[\ell,k \right],
\end{align}
where $\vartheta_{s,u} = \alpha_{s,u}G(\theta_{s,u}^{\mathrm{el}})$, $\gamma[\ell,k] = e^{\jmath 2 \pi(\ell T\upsilon_{s,u}-k\Delta f\tau_{s,u})}$, $\mathbf{f}_s[\ell,k] = \mathbf{F}_s\mathbf{t}_s[\ell,k]$ with $\mathbf{F}_s \in \mathbb{C}^{N \times U}$ being the beamforming matrix, and $z_u[\ell,k]\sim \mathcal{CN}(0,N_0 \Delta f )$ is the \ac{awgn} with single-side \ac{psd}~$N_0$.
Without loss of generality, the following derivation focuses on the $u$-th \ac{ut}.

Let ${\bm{\eta}}_{s,u} =[\acute{\bm{\eta}}_{s,u}^\mathsf{T},\grave{\bm{\eta}}_{s,u}^\mathsf{T}]^\mathsf{T} \in \mathbb{R}^{6}$ denote the unknown channel-domain parameters in the signal received from the $s$-th satellite, where $\acute{\bm{\eta}}_{s,u} = [\upsilon_{s,u},\tau_{s,u}]^\mathsf{T} \in \mathbb{R}^{2}$ are the parameters used for \ac{ut} positioning, and $\grave{\bm{\eta}}_{s,u} = [\bm{\theta}_{s,u}^\mathsf{T},\Re (\vartheta_{s,u}),\Im (\vartheta_{s,u})]^\mathsf{T} \in \mathbb{R}^{4}$ are treated as nuisance parameters. Delay $\tau_{s,u}$ and Doppler $\upsilon_{s,u}$ are directly determined by the UT's position and velocity relative to the satellites, making them the most reliable observables for positioning. In contrast, the \ac{aod} $\bm{\theta}_{s,u}$ is not exploited here due to the limited angular resolution of \ac{leo} satellites operating at high altitudes, which makes \ac{aod} estimates unreliable unless extremely large arrays are employed. Similarly, the channel gain $\vartheta_{s,u}$ is heavily influenced by environmental variability and hardware impairments and thus provides little stable geometric information for positioning, except in fingerprinting-based approaches~\cite{henk2022cl_i}.

We further concatenate all the utilized channel-domain parameters collected from $S$ satellites as $\acute{\bm{\eta}}_u = [\acute{\bm{\eta}}_{1,u}^\mathsf{T},\ldots,\acute{\bm{\eta}}_{S,u}^\mathsf{T}]^\mathsf{T} \in \mathbb{R}^{2S}$.
Therefore, in the large-timescale downlink positioning, we first estimate $\acute{\bm{\eta}}_u$ based on $y_{s,u}[\ell,k]$, $s=1,\dots,S$, $\ell=1,\dots,L_{\text{p}}$, $k=1,\dots,K$, where $L_{\text{p}}$ denotes the number of pilot symbols for downlink positioning. Afterwards, we estimate the $u$-th \ac{ut}'s position, clock biases, and \acp{cfo}. 

\subsubsection{The FIM of \texorpdfstring{${\bm{\eta}}_{s,u}$}{TEXT1}}
Let ${\mu}_{s,u}[\ell,k]$ denote the noise-free version of $y_{s,u}[\ell,k]$ and $\mathbf{G}_{s,u}[\ell,k] = [\mathbf{g}_v[\ell,k],\mathbf{g}_\tau[\ell,k],\mathbf{g}_{\theta_{\mathrm{az}}}[\ell,k],\mathbf{g}_{\theta_{\mathrm{el}}}[\ell,k],\mathbf{g}_{\vartheta,\mathrm{R}}[\ell,k],\mathbf{g}_{\vartheta,\mathrm{I}}[\ell,k]] \in \mathbb{C}^{N \times 6}$. Here, the columns fulfill 
\begin{subequations}
\begin{align}
    \mathbf{g}_v\left[\ell,k \right] &= \vartheta_{s,u} \gamma\left[\ell,k\right] \left(\jmath 2 \pi \ell T\right)\mathbf{a}\left(\bm{\theta}_{s,u}\right),\\
    \mathbf{g}_\tau\left[\ell,k \right] &= \vartheta_{s,u}  \gamma\left[\ell,k\right] \left(-\jmath 2 \pi k\Delta f\right)\mathbf{a}\left(\bm{\theta}_{s,u}\right),\\
    \mathbf{g}_{\theta_{\mathrm{az}}}\left[\ell,k \right] &= \vartheta_{s,u}  \gamma\left[\ell,k\right]\frac{\partial \mathbf{a}\left(\bm{\theta}_{s,u}\right)}{\partial \theta_{s,u}^{\mathrm{az}}},\\
    \mathbf{g}_{\theta_{\mathrm{el}}}\left[\ell,k \right] 
     &= \vartheta_{s,u}\gamma\left[\ell,k\right] \frac{\partial \mathbf{a}\left(\bm{\theta}_{s,u}\right)}{\partial \theta_{s,u}^{\mathrm{el}}},\\
    \mathbf{g}_{\vartheta,\mathrm{R}}\left[\ell,k \right] &=  \gamma\left[\ell,k\right] \mathbf{a}\left(\bm{\theta}_{s,u}\right),\\
    \mathbf{g}_{\vartheta,\mathrm{I}}\left[\ell,k \right] &= \jmath  \gamma\left[\ell,k\right] \mathbf{a}\left(\bm{\theta}_{s,u}\right).    
\end{align} 
\end{subequations}
Based on the signal received from the $s$-th satellite, the \ac{fim} of ${\bm{\eta}}_{s,u}$ can be computed by the Slepian-Bangs formula as
\begin{align}
    &\mathbf{J}_s({\bm{\eta}}_{s,u})\notag\\
    =& \frac{2}{N_0 \Delta f}\sum_{\ell=1}^{L_{\text{p}}}\sum_{k=1}^K\Re \Big\{\Big(\frac{\partial {\mu}_{s,u}\left[\ell,k \right]}{\partial {\bm{\eta}}_{s,u}}\Big)\Big(\frac{\partial {\mu}_{s,u}\left[\ell,k \right]}{\partial {\bm{\eta}}_{s,u}}\Big)^\mathsf{H}\Big\}, \\
    =& \frac{2}{N_0 \Delta f}\sum_{\ell=1}^{L_{\text{p}}}\sum_{k=1}^K\Re \big\{\mathbf{G}_{s,u}^\mathsf{T}\left[\ell,k \right]\mathbf{f}_s\left[\ell,k\right]\mathbf{f}_s\left[\ell,k\right]^\mathsf{H}\mathbf{G}_{s,u}^{*}\left[\ell,k \right]\big\}.\notag
\end{align}

\subsubsection{The FIM of $\acute{\bm{\eta}}_{u}$}\label{sec:FIM_etau}
Partitioning $\mathbf{J}_s(\bm{\eta}_{s,u}) = [ \mathbf{X}, \mathbf{Y}; \mathbf{Y}^\mathsf{T}
, \mathbf{Z}]$,
where $\mathbf{X} \in \mathbb{R}^{2\times 2}$, we can compute the \ac{fim} of the total utilized channel-domain parameters $\acute{\bm{\eta}}_{s,u}$ as $\mathbf{J}_s\left(\acute{\bm{\eta}}_{s,u}\right) = \mathbf{X} - \mathbf{Y} \mathbf{Z}^{-1} \mathbf{Y}^{\mathsf{T}}$. Then, we can obtain the \ac{fim} of $\acute{\bm{\eta}}_u$ as
\begin{equation}
    \mathbf{J}\left(\acute{\bm{\eta}}_{u}\right) = \text{blkdiag}\big\{\mathbf{J}_1\left(\acute{\bm{\eta}}_{1,u}\right),\ldots,\mathbf{J}_S\left(\acute{\bm{\eta}}_{S,u}\right)\big\}.
\end{equation}

Based on this, we can infer that the achievable lowest estimation error on $\acute{\bm{\eta}}_{u}$ regardless of the adopted algorithms is characterized by the Gaussian distribution $\mathcal{N}\big(\mathbf{0},\mathbf{J}^{-1}(\acute{\bm{\eta}}_u)\big)$. Any algorithm achieving this error covariance matrix is called an \textit{efficient estimator}.

\subsubsection{MCRB and LB Derivation}
Suppose an efficient estimator is utilized to estimate $\acute{\bm{\eta}}_u$. We now focus on estimating the position-domain parameters, where the aforementioned model mismatches occur. We consider a scenario where each satellite-\ac{ut} pair has a distinct clock bias $b_{s,u}$ and \ac{cfo} $\delta_{s,u}$. For convenience, we use $b_u$ and $\delta_u$ to denote an average clock bias and \ac{cfo} among the $S$ satellites, thus we can rewrite these individual clock biases and \acp{cfo} as $b_{s,u}=b_u + \Delta b_{s,u}$ and $\delta_{s,u} = \delta_u + \Delta\delta_{s,u}$, respectively, where $\Delta b_{s,u}$ and $\Delta\delta_{s,u}$ denote the differences on each clock bias and \ac{cfo}. 

Let $\mathbf{v}_s \in \mathbb{R}^{3}$ be the velocity of the $s$-th \ac{leo} satellite. Here, as the velocity of \ac{leo} satellite is usually much higher than that of the \ac{ut}, we ignore the velocity contribution from the \ac{ut} side. This assumption is justified since \ac{leo} satellites move at around 7.6 km/s, whereas \acp{ut} typically move at only a few m/s (pedestrians) to a few hundred m/s (vehicles/aircraft), making their velocity contribution two to three orders of magnitude smaller and thus negligible for the Doppler term.
Based on the underlying geometric relationship, the forward model between the entries of $\acute{\bm{\eta}}_u$ and unknown position of the $u$-th \ac{ut} can be expressed as
\begin{equation}\label{eq:TM-doppler}
\upsilon_{s,u} = \frac{\mathbf{v}_s^\mathrm{T}(\mathbf{p}_u-\mathbf{p}_s)}{\lambda\|\mathbf{p}_u-\mathbf{p}_s\|} + \delta_u + \Delta\delta_{s,u}   
\end{equation}
and
\begin{equation}\label{eq:TM-delay}
\tau_{s,u} = \frac{\|\mathbf{p}_u-\mathbf{p}_s\|}{c} + b_{u} + \Delta b_{s,u},    
\end{equation}
which we refer to as the \textit{true model}. Here, $\mathbf{p}_u \in \mathbb{R}^3$ and $\mathbf{p}_s \in \mathbb{R}^3$ represent the positions of the $u$-th \ac{ut} and the $s$-th satellite\footnote{Note that \ac{leo} satellites' positions are assumed precisely known at the \acp{ut}. This information can be obtained from the ephemerides contained in the two-line element files, which are updated daily by the North American Aerospace Defense Command (NORAD)~\cite{pinjun2024jsac}.}, respectively. 

In real applications, we often ignore these clock bias and \ac{cfo} differences, i.e., assuming all satellite-\ac{ut} pairs have the same clock bias and \ac{cfo}, for algorithmic simplicity. To account for the impact of such model mismatch, we derive the \ac{mcrb}, with its fundamentals detailed in Appendix~\ref{mcrb_fundamental} for completeness. When neglecting the difference on each clock bias and \ac{cfo}, the \textit{mismatched model} is given by
\begin{equation}\label{eq:MM-Doppler}
\upsilon_{s,u} = \frac{\mathbf{v}_s^\mathrm{T}(\mathbf{p}_u-\mathbf{p}_s)}{\lambda\|\mathbf{p}_u-\mathbf{p}_s\|} + \delta_u   
\end{equation}
and
\begin{equation}\label{eq:MM-delay}
\tau_{s,u} = \frac{\|\mathbf{p}_u-\mathbf{p}_s\|}{c} + b_{u}.    
\end{equation}

Now, we suppose an estimator that aims to estimate the unknown position of the $u$-th \ac{ut} based on the realistic observation $\acute{\bm{\eta}}_u$ and mismatched model, the total unknown parameters in the mismatch model can be concatenated as $\mathbf{r}_u = [\mathbf{p}_u^\mathsf{T},\delta_u,b_u]^\mathsf{T}\in \mathbb{R}^{5}$. Based on the true model and mismatched model, we can respectively express the true likelihood function $\xi_\mathrm{T}$ and the mismatched likelihood function $\xi_\mathrm{M}$ as 
\begin{equation}
{\xi_\mathrm{T}}\left(\acute{\bm{\eta}}_u;\mathbf{r}_u\right)\propto e^{\big(\acute{\bm{\eta}}_u-\bar{\mathbf{f}}\left(\mathbf{r}_u\right)\big)^\mathsf{T}\bm{\Sigma}^{-1}\big(\acute{\bm{\eta}}_u-\bar{\mathbf{f}}\left(\mathbf{r}_u\right)\big) }  
\end{equation}
and
\begin{equation}
{\xi_\mathrm{M}}\left(\acute{\bm{\eta}}_u;\mathbf{r}_u\right)\propto
e^{\big(\acute{\bm{\eta}}_u-\tilde{\mathbf{f}}\left(\mathbf{r}_u\right)\big)^\mathsf{T}\bm{\Sigma}^{-1}\big(\acute{\bm{\eta}}_u-\tilde{\mathbf{f}}\left(\mathbf{r}_u\right)\big)},
\end{equation}
respectively, where the function $\bar{\mathbf{f}}(\cdot)$ maps $\mathbf{r}_u$ to $\acute{\bm{\eta}}_u$ according to true model, and the function $\tilde{\mathbf{f}}(\cdot)$ maps $\mathbf{r}_u$ to $\acute{\bm{\eta}}_u$ according to mismatched model. As mentioned in Section~\ref{sec:FIM_etau}, $\acute{\bm{\eta}}_u$ is the estimated channel-domain parameters through an efficient estimator, i.e., $\acute{\bm{\eta}}_u\sim\mathcal{N}(\bar{\mathbf{f}}(\mathbf{r}_u),\bm{\Sigma})$, where $\bm{\Sigma}=\mathbf{J}^{-1}(\acute{\bm{\eta}}_u)$.

\begin{itemize}
\item \emph{Pseudo-True Parameter:} 
According to \eqref{pseud_para}, the pseudo-true parameter, denoted by $\tilde{\mathbf{r}}_u$, is given by
\begin{align}\label{pos_pesudo_true}
    \tilde{\mathbf{r}}_u &=\arg\min_{\mathbf{r}_u}\ \mathcal{D}\big\{{\xi_\mathrm{T}}(\acute{\bm{\eta}}_u;\bar{\mathbf{r}}_u)\|{\xi_\mathrm{M}}(\acute{\bm{\eta}}_u;\mathbf{r}_u)\big\},\\
    &=\arg\min_{\mathbf{r}_u}\ \big(\bar{\mathbf{f}}(\bar{\mathbf{r}}_u) - \tilde{\mathbf{f}}\left(\mathbf{r}_u\right)\big)^\mathsf{T} \bm{\Sigma}^{-1} \big(\bar{\mathbf{f}}(\bar{\mathbf{r}}_u) - \tilde{\mathbf{f}}\left(\mathbf{r}_u\right)\big),\notag
\end{align}
where the top bar is used to highlight the ground-truth position-domain parameter and forward model. 
The above problem can be solved using gradient descent with backtracking line search, wherein the true parameters serve as the initial point\cite{cuneyd2024twc}. 

\item \emph{MCRB:} 
According to Appendix \ref{mcrb_fundamental}, the generalized \acp{fim}, i.e., $\mathbf{A}_{\tilde{\mathbf{r}}_u}$ and $\mathbf{B}_{\tilde{\mathbf{r}}_u}$, can be computed by determining the first- and second-order derivatives of $\tilde{\mathbf{f}}\left(\mathbf{r}_u\right)$ with respect to the elements of $\mathbf{r}_u$, as detailed in Appendix \ref{deriv_pos}. Subsequently, the \ac{mcrb} is expressed as 
\begin{equation}
\text{MCRB}\left(\tilde{\mathbf{r}}_u\right) = \mathbf{A}_{\tilde{\mathbf{r}}_u}^{-1}\mathbf{B}_{\tilde{\mathbf{r}}_u}\mathbf{A}_{\tilde{\mathbf{r}}_u}^{-1}.
\end{equation}

\item \emph{LB:}
Finally, the estimation error \ac{lb} matrix can be computed as
\begin{equation}\label{eq:LBM}
    \text{LBM}\left(\tilde{\mathbf{r}}_u\right) = \text{MCRB}\left(\tilde{\mathbf{r}}_u\right) + \text{Bias}\left(\tilde{\mathbf{r}}_u\right),
\end{equation}
where $\text{Bias}\left(\tilde{\mathbf{r}}_u\right) = \left(\bar{\mathbf{r}}_u-\tilde{\mathbf{r}}_u\right)\left(\bar{\mathbf{r}}_u-\tilde{\mathbf{r}}_u\right)^\mathsf{T}$.
Based on~\eqref{eq:LBM}, the \ac{lb} for the expected \ac{rmse} of the $u$-th \ac{ut}'s position estimation in the presence of model mismatch is given by
\begin{equation}
\text{LB}\left(\mathbf{p}_u\right) = \sqrt{\text{Tr}\big(\left[\text{LBM}\left(\tilde{\mathbf{r}}_u\right)\right]_{1:3,1:3}\big)}.
\end{equation} 
\end{itemize}

\subsection{Small-Timescale Uplink Channel Estimation}
\subsubsection{Signal Model}
After obtaining the positions of \acp{ut}, these large-timescale parameters are fed back to the \ac{leo} satellites. The second phase of \ac{pace} then focuses on estimating the small-timescale parameters (i.e., the channel gains) at the \ac{leo} satellites through uplink channel estimation, followed by complete channel reconstruction.

Let $\mathbf{d}_{s,u}[\ell,k] = \mathbf{a}(\boldsymbol{\theta}_{s,u})\sqrt{P_u/K} t_u\left[\ell,k\right]$, where $P_u$ denotes the transmit power of the $u$-th \ac{ut}, and $t_u[\ell,k]$ represents the unit-modulus pilot transmitted by the $u$-th \ac{ut} during the $\ell$-th symbol and the $k$-th subcarrier. Uniform power allocation is employed across all subcarriers. As previously mentioned, the Doppler shift can be largely mitigated in \ac{leo} satellite communications. After being exploited in downlink positioning, it is excluded from the signal model including uplink channel estimation and downlink communication for simplicity. Here, we also assume \acp{ut} are multiplexed orthogonally using \ac{tdma}. By channel reciprocity, the signal received by the $s$-th \ac{leo} satellite from the $u$-th \ac{ut} is expressed as  
\begin{equation}\label{up_ce_true}
    \mathbf{y}_{s,u}\left[\ell,k \right] = \underbrace{\vartheta_{s,u} e^{-\jmath 2 \pi k \Delta f \tau_{s,u}} \mathbf{d}_{s,u}[\ell,k]}_{\bar{\mathbf{g}}_{s,u}\left[\ell,k \right]} + \mathbf{z}_s\left[\ell,k \right],
\end{equation}
where $\mathbf{z}_s[\ell,k]\sim \mathcal{CN}(\mathbf{0},N_0 \Delta f \mathbf{I}_N)$ represents the \ac{awgn}. 

We observe that $\mathbf{d}_{s,u}[\ell,k]$ is directly determined by the \acp{ut}' position feedback. Consequently, instead of estimating the full channel vector, it suffices to estimate only two scalars, i.e., the channel gain $\vartheta_{s,u}$ and the \ac{toa} $\tau_{s,u}$.\footnote{Note that although the \ac{toa} $\tau_{s,u}$ has already been estimated during the downlink positioning process, it will nevertheless be estimated during uplink channel estimation as part of standard \ac{ofdm} processing~\cite{simoCE2008cl}. This is necessary unless the \ac{ut} and the \ac{leo} satellite are perfectly time-synchronized and the propagation delay remains precisely known, which are hardly satisfied in the highly dynamic \ac{leo} satellite environment.} This significantly simplifies the uplink channel estimation task. However, since the position information is based on estimation rather than ground truth, model mismatch arises again. The received signal based on the estimated position information (mismatched model) is expressed as
\begin{equation}\label{up_ce_mis}
    \mathbf{y}_{s,u}\left[\ell,k \right] = \underbrace{\vartheta_{s,u} e^{-\jmath 2 \pi k \Delta f \tau_{s,u}} \tilde{\mathbf{d}}_{s,u}\left[\ell,k \right]}_{\tilde{\mathbf{g}}_{s,u}\left[\ell,k \right]}  + \mathbf{z}_s\left[\ell,k \right],
\end{equation}
where $\tilde{\mathbf{d}}_{s,u}[\ell,k] = \mathbf{a}(\tilde{\boldsymbol{\theta}}_{s,u})\sqrt{P_u/K} t_u\left[\ell,k\right]$. Here, $\tilde{\boldsymbol{\theta}}_{s,u}$ represents the calculated \ac{aoa} based on the estimated $\tilde{\mathbf{p}}_u$ and the geometrical relationship, given by
\begin{subequations}
    \begin{align}
    \tilde{\theta}_{s,u}^{\mathrm{az}} &= \text{atan2}\left(\left[\tilde{\mathbf{p}}_u - \mathbf{p}_s\right]_2,\left[\tilde{\mathbf{p}}_u - \mathbf{p}_s\right]_1\right), \\
    \tilde{\theta}_{s,u}^{\mathrm{el}} &= \text{asin}\Big(\frac{\left[\left(\tilde{\mathbf{p}}_u - \mathbf{p}_s\right)\right]_3}{\left\|\tilde{\mathbf{p}}_u - \mathbf{p}_s\right\|}\Big).
    \end{align}
\end{subequations}

\subsubsection{MCRB and LB Derivation}
We denote the total temporal and spectral observation as $\mathbf{y}_{s,u} = [\mathbf{y}_{s,u}^{\mathsf{T}}[1,1], \ldots, \mathbf{y}_{s,u}^{\mathsf{T}}[1,K],\ldots,\mathbf{y}_{s,u}^{\mathsf{T}}[L_{\text{c}},K]]^{\mathsf{T}} \in \mathbb{C}^{L_{\text{c}} K N}$, where $L_{\text{c}}$ represents the number of pilot symbols used for uplink channel estimation. Let $\boldsymbol{\eta}_{s,u} = [\Re (\vartheta_{s,u}),\Im (\vartheta_{s,u}), \tau_{s,u}]^{\mathsf{T}} \in \mathbb{R}^3$ collect the parameters to be estimated in the uplink channel estimation. The likelihood functions under the true and mismatched models are given by  
\begin{equation}
{\xi_\mathrm{T}}\left(\mathbf{y}_{s,u};\boldsymbol{\eta}_{s,u}\right)\propto e^{\left\|\mathbf{y}_{s,u}-\bar{\mathbf{g}}\left(\boldsymbol{\eta}_{s,u}\right)\right\|^2}  
\end{equation}  
and  
\begin{equation}
{\xi_\mathrm{M}}\left(\mathbf{y}_{s,u};\boldsymbol{\eta}_{s,u}\right)\propto e^{\left\|\mathbf{y}_{s,u}-\tilde{\mathbf{g}}\left(\boldsymbol{\eta}_{s,u}\right)\right\|^2},  
\end{equation}  
respectively, where we define $\bar{\mathbf{g}}(\boldsymbol{\eta}_{s,u}) =  [\bar{\mathbf{g}}_{s,u}^{\mathsf{T}}[1,1], \ldots,\bar{\mathbf{g}}_{s,u}^{\mathsf{T}}[1,K],\ldots,\bar{\mathbf{g}}_{s,u}^{\mathsf{T}}[L_{\text{c}},K]]^{\mathsf{T}}  \in \mathbb{C}^{L_{\text{c}} K N}$ and $\tilde{\mathbf{g}}(\boldsymbol{\eta}_{s,u}) = [\tilde{\mathbf{g}}_{s,u}^{\mathsf{T}}[1,1], \ldots,\tilde{\mathbf{g}}_{s,u}^{\mathsf{T}}[1,K],\ldots,\tilde{\mathbf{g}}_{s,u}^{\mathsf{T}}[L_{\text{c}},K]]^{\mathsf{T}} \in \mathbb{C}^{L_{\text{c}} K N}$.

\begin{itemize}
\item \emph{Pseudo-True Parameter:}
According to Appendix \ref{mcrb_fundamental}, the pseudo-true parameter, denoted by $\tilde{\boldsymbol{\eta}}_{s,u}$, is derived as  
\begin{align}  
    \tilde{\boldsymbol{\eta}}_{s,u} &= \arg\min_{\boldsymbol{\eta}_{s,u}} \ \mathcal{D}\big\{{\xi_\mathrm{T}}\left(\mathbf{y}_{s,u};\bar{\boldsymbol{\eta}}_{s,u}\right)\|{\xi_\mathrm{M}}\left(\mathbf{y}_{s,u};\boldsymbol{\eta}_{s,u}\right)\big\}, \notag  \\  
    &= \arg\min_{\boldsymbol{\eta}_{s,u}} \ \left\|\bar{\mathbf{g}}\left(\bar{\boldsymbol{\eta}}_{s,u}\right) - \tilde{\mathbf{g}}\left(\boldsymbol{\eta}_{s,u}\right)\right\|^2,  
\end{align}  
where the bar above the parameter emphasizes the ground-truth value. Once again, by employing gradient descent and treating truth parameter as initial point, the above problem can be efficiently solved \cite{cuneyd2024twc}.  

\item \emph{MCRB:}  
The generalized \acp{fim}, i.e., $\mathbf{A}_{\tilde{\boldsymbol{\eta}}_{s,u}}$ and $\mathbf{B}_{\tilde{\boldsymbol{\eta}}_{s,u}}$, are obtained by calculating the first- and second-order derivatives of $\tilde{\mathbf{q}}\left(\boldsymbol{\eta}_{s,u}\right)$ with respect to the elements of $\boldsymbol{\eta}_{s,u}$, as described in Appendix \ref{deriv_chan}. Based on these, the \ac{mcrb} is given by  
\begin{equation}
\text{MCRB}\left(\tilde{\boldsymbol{\eta}}_{s,u}\right) = \mathbf{A}_{\tilde{\boldsymbol{\eta}}_{s,u}}^{-1}\mathbf{B}_{\tilde{\boldsymbol{\eta}}_{s,u}}\mathbf{A}_{\tilde{\boldsymbol{\eta}}_{s,u}}^{-1}.
\end{equation}

\item \emph{LB:}  
The estimation error \ac{lb} matrix is then calculated as  
\begin{equation}\label{lbm_chan}
    \text{LBM}\left(\tilde{\boldsymbol{\eta}}_{s,u}\right) = \text{MCRB}\left(\tilde{\boldsymbol{\eta}}_{s,u}\right) + \text{Bias}\left(\tilde{\boldsymbol{\eta}}_{s,u}\right),
\end{equation}  
where $\text{Bias}\left(\tilde{\boldsymbol{\eta}}_{s,u}\right) = \left(\bar{\boldsymbol{\eta}}_{s,u} - \tilde{\boldsymbol{\eta}}_{s,u}\right)\left(\bar{\boldsymbol{\eta}}_{s,u} - \tilde{\boldsymbol{\eta}}_{s,u}\right)^\mathsf{T}$.
For an efficient estimator, the \ac{lb} for the expected \ac{rmse} of the channel gain $\vartheta_{s,u}$ and the \ac{toa} $\tau_{s,u}$, in the presence of model mismatch, are expressed as  
\begin{equation}
\text{LB}\left(\vartheta_{s,u}\right) = \sqrt{\text{Tr}\big(\left[\text{LBM}\left(\tilde{\boldsymbol{\eta}}_{s,u}\right)\right]_{1:2,1:2}\big)}
\end{equation}  
and  
\begin{equation}
\text{LB}\left(\tau_{s,u}\right) = \sqrt{\left[\text{LBM}\left(\tilde{\boldsymbol{\eta}}_{s,u}\right)\right]_{3,3}},
\end{equation}  
respectively.
\end{itemize}

\section{Cooperative Beamforming for Downlink Communications}\label{sec-coop}
The proposed two-timescale \ac{pace} enables channel reconstruction between each \ac{leo} satellite and all \acp{ut}, enabling \ac{csi}-based beamforming for downlink communication. To address the limited link budget in single-satellite downlink scenarios, we introduce multi-\ac{leo} satellite cooperative beamforming, where each \ac{ut} is served by the satellites within the cluster simultaneously. 
Note that the cooperative beamforming optimization developed in this section is based on a \emph{nominal} signal model, which treats the channel parameters estimated via \ac{pace} as true and assumes perfect delay/Doppler compensation across satellites. In contrast, the communication rate in the simulations is evaluated using the \emph{ground-truth} channel parameters under imperfect compensation. \emph{This separation enables a direct assessment of the effectiveness of \ac{pace} by examining the performance achieved by beamformers optimized with nominal (\ac{pace}-estimated) parameters.} The smaller the performance gap between nominal- and ground-truth-optimized beamformers, the more reliable \ac{pace} can be considered. Developing robust beamforming strategies that explicitly account for \ac{csi} uncertainty (e.g.,~\cite{fanjiang2023twc,zack2023twc,zack2024tcom}) and/or asynchronous interference due to imperfect compensation (e.g.,~\cite{asyn2025tvt}) represents an important but orthogonal research direction, which is left for future work.

\subsection{Nominal Signal Model}
In the nominal signal model, we assume perfect delay and Doppler compensation across all satellites. Under this assumption, the nominal channel, denoted by $\tilde{\mathbf{h}}_{s,u} = \tilde{\vartheta}_{s,u}\mathbf{a}(\tilde{\boldsymbol{\theta}}_{s,u})$, is frequency-flat, where the tilde denotes estimated values. This allows subcarrier-coherent beamforming, where the same beamformer is applied across all subcarriers, thereby reducing optimization complexity. Specifically, for the $s$-th \ac{leo} satellite transmitting over the $k$-th subcarrier, the signal is
\begin{equation}
\mathbf{x}_s \left[k\right] = \mathbf{W}_s \mathbf{s}\left[k\right],   
\end{equation}
where $\mathbf{W}_s = [\mathbf{w}_{s,1},\mathbf{w}_{s,2},\ldots,\mathbf{w}_{s,U}] \in \mathbb{C}^{N \times U}$ represents the beamformer for the $s$-th \ac{leo} satellite, and $\mathbf{s}\left[k\right] \sim \mathcal{CN}(\mathbf{0},\mathbf{I}_U)$ denotes the data streams transmitted to the $U$ \acp{ut} over the $k$-th subcarrier.  
The received signal at the $u$-th \ac{ut} over the $k$-th subcarrier is then
\begin{align}
y_u\left[k\right] = \sum_{s=1}^{S}\tilde{\mathbf{h}}_{s,u}^{\mathsf{T}} \mathbf{w}_{s,u}  s_u \left[k\right] + \underbrace{\sum_{m \neq u}^{U} \sum_{s=1}^{S} \tilde{\mathbf{h}}_{s,u}^{\mathsf{T}} \mathbf{w}_{s,m} s_m \left[k\right]}_{\text{Inter-user interference}}+ z_u \left[k\right],
\end{align}
where $z_u [k]\sim \mathcal{CN}(0,N_0 \Delta f )$ denotes the \ac{awgn}. 

The nominal achievable communication rate for the $u$-th \ac{ut}, aggregated across all subcarriers, is given by
\begin{equation}
\tilde{R}_{u}  = B \log_2 \left(1 + \frac{\left|\sum \limits_{s=1}^{S}\tilde{\mathbf{h}}_{s,u}^{\mathsf{T}} \mathbf{w}_{s,u}\right|^2}{\sum\limits_{m \neq u}^{U} \left|\sum\limits_{s=1}^{S}\tilde{\mathbf{h}}_{s,u}^{\mathsf{T}} \mathbf{w}_{s,m}\right|^2+ N_0 \Delta f}\right). 
\end{equation}

\subsection{Problem Formulation}
We aim to optimize the beamformers at each \ac{leo} satellite to maximize the sum rate of all \acp{ut}. The optimization problem is formulated as follows 
\begin{align}\label{ori_prob}
\mathop {\max }\limits_{\mathbf{W}_s} \;\;\; & \sum_{u=1}^{U} \tilde{R}_{u} \notag \\
{\rm{s.t.}}\;\;\;
& \sum_{u=1}^{U} \left\|\mathbf{w}_{s,u}\right\|^2 \le \frac{P_s}{K}, \;\forall s,
\end{align}
where $P_s$ denotes the power budget at the $s$-th \ac{leo} satellite. 

\renewcommand{\algorithmicrequire}{\textbf{Input:}}
\renewcommand{\algorithmicensure}{\textbf{Output:}}
\begin{algorithm}[t]
\caption{Proposed Algorithm for Solving \eqref{ori_prob}}
\label{wmmse_bf_opt}
\begin{algorithmic}[1]
\State \textbf{Initialize}: {$\mathbf{W}_s,\forall s$;}
\Repeat
\State {Update $\mu_u$ using \eqref{update_mu};}
\State {Update $\omega_{u}$ using \eqref{update_omega};}
\For {$s = 1 : S$}
\State {Obtain eigenvalue decomposition of $\mathbf{R}_s$;}
\State {Determine $\lambda$ by Golden-section search over \eqref{final_Search_form};}
\State {Determine $\mathbf{w}_s$ via \eqref{optimal_bf_form_lc} and reshape to obtain $\mathbf{W}_s$;}
\EndFor
\Until {the reduction ratio of the objective value falls below a predefined threshold or a predetermined iteration number is reached;}
\State \textbf{Output}: {$\mathbf{W}_s,\forall s$.}
\end{algorithmic}
\end{algorithm}

\subsection{Proposed Cooperative Beamforming Approach}
In the following, we employ the \ac{wmmse} optimization framework \cite{shi2011wmmse} to solve \eqref{ori_prob} iteratively. Specifically, by introducing the auxiliary variables $\mu_u$ and $\omega_{u}$, where $\mu_u$ and $\omega_{u}$ serve as the MMSE equalizer and weight, respectively \cite{shi2011wmmse}, \eqref{ori_prob} can be equivalently reformulated as
\begin{align}\label{dist_prob_refor}  
\mathop {\max }\limits_{\mu_u,\omega_{u},\mathbf{W}_s} \;\;\; &\sum_{u=1}^{U} \left(\ln \omega_{u} - \omega_{u} \Psi_{u}\right) \notag \\  
{\rm{s.t.}}\;\;\;  
& \sum_{u=1}^{U} \left\|\mathbf{w}_{s,u}\right\|^2 \le \frac{P_s}{K}, \;\forall s, 
\end{align}  
where $\Psi_{u} = |1 - \mu_u\sum_{s=1}^{S}\tilde{\mathbf{h}}_{s,u}^{\mathsf{T}} \mathbf{w}_{s,u} |^2 + |\mu_u|^2 (\sum_{m \neq u}^{U} |\sum_{s=1}^{S}\tilde{\mathbf{h}}_{s,u}^{\mathsf{T}} \mathbf{w}_{s,m}|^2 + N_0 \Delta f)$. The optimization problem in \eqref{dist_prob_refor} is solved by iteratively updating $\mathbf{W}_s$, $\omega_{u}$, and $\mu_u$, as detailed below.

\subsubsection{Update $\mu_u$}
With $\omega_{u}$ and $\mathbf{W}_s$ fixed, the optimal $\mu_u$ is derived by setting $\partial \Psi_{u} / \partial \mu_u = 0$, resulting in
\begin{align}\label{update_mu}  
\mu_u =\frac{\left(\sum_{s=1}^{S}\tilde{\mathbf{h}}_{s,u}^{\mathsf{T}} \mathbf{w}_{s,u}\right)^*}{ \sum_{m =1}^{U} \left|\sum_{s=1}^{S}\tilde{\mathbf{h}}_{s,u}^{\mathsf{T}} \mathbf{w}_{s,m}\right|^2 + N_0 \Delta f}.
\end{align}

\subsubsection{Update $\omega_{u}$}
With $\mu_u$ and $\mathbf{W}_s$ fixed, the optimal $\omega_{u}$ is obtained by setting $\partial \Psi_{u} / \partial \omega_{u} = 0$, leading to 
\begin{equation}\label{update_omega}  
\omega_{u} = \frac{1}{\Psi_{u}}.  
\end{equation}

\subsubsection{Update $\mathbf{W}_s$}\label{update_bf} 
With $\mu_u$ and $\omega_{u}$ fixed, the optimization problem to find the optimal $\mathbf{W}_s$ is formulated as  
\begin{align}\label{dist_prob_bf_opt}  
\mathop {\min }\limits_{\mathbf{W}_s} \;\;\; &\sum_{u=1}^{U} \omega_{u}\Psi_{u} \notag \\  
{\rm{s.t.}}\;\;\;  
& \sum_{u=1}^{U} \left\|\mathbf{w}_{s,u}\right\|^2 \le \frac{P_s}{K}, \;\forall s.  
\end{align}  
Here, $\Psi_{u}$ can be expressed as the sum of a positive semi-definite quadratic form and a linear term. 

We note that \eqref{dist_prob_bf_opt} is a convex \ac{qcqp} with respect to $\mathbf{W}_s$, which can be solved using standard toolboxes such as CVX. However, this approach incurs a per-iteration complexity of $\mathcal{O}((SNU)^3)$, since the beamformers of all satellites are optimized jointly. To better suit computation resource-constrained \ac{leo} scenarios, we instead develop a low-complexity method that strategically exploits \ac{bcd}, strong duality, and matrix inversion properties.\footnote{Despite the significant complexity reduction enabled by the proposed scheme, we note that the optimization is still carried out in a centralized manner (e.g., at a cluster-head satellite), which may incur scalability bottlenecks as the network size grows. Designing decentralized strategies that distribute the optimization across participating satellites to improve scalability is an important research direction but lies beyond the scope of this paper. We refer interested readers to our recent work~\cite{zack2025twc} for a scalable distributed beamforming framework over networked LEO satellites.} This reduces the dominant complexity of solving \eqref{dist_prob_bf_opt} from $\mathcal{O}((SNU)^3)$ to $\mathcal{O}(SN^3)$. Further details are provided in Appendix~\ref{lc_algo}. The overall procedure for solving \eqref{ori_prob} is summarized in Algorithm~\ref{wmmse_bf_opt}, and its convergence follows directly from the \ac{bcd} structure of the algorithm.

\newcounter{MYtempeqncn}
\begin{figure*}[!b]
\normalsize
\setcounter{MYtempeqncn}{\value{equation}}
\setcounter{equation}{37}
\hrulefill
\begin{equation}\label{actual_rate_model}
R_{u}
= \Delta f \sum_{k=1}^{K}\log_2\!\left(1 +
\frac{\left|\sum \limits_{s=1}^{S}e^{-\jmath 2 \pi k\Delta f\left(\tau_{s,u} - \tilde{\tau}_{s,u}\right)}\bar{\mathbf{h}}_{s,u}^{\mathsf{T}}\!\left[k\right] \mathbf{w}_{s,u}\right|^2}{
\sum\limits_{m \neq u}^{U}
\left|\sum\limits_{s=1}^{S}e^{-\jmath 2 \pi k\Delta f\left(\tau_{s,u} - \tilde{\tau}_{s,m}\right)}\bar{\mathbf{h}}_{s,u}^{\mathsf{T}}\!\left[k\right]\mathbf{w}_{s,m}\right|^2
+ N_0 \Delta f}\right),
\end{equation}
\setcounter{equation}{\value{MYtempeqncn}}
\end{figure*}

\subsection{Evaluation Model}\label{sec:eval_model}
As stated above, after obtaining the cooperative beamformers by solving the optimization problem under the \emph{nominal} signal model (with \ac{pace}-estimated channel parameters and ideal inter-satellite compensation), we evaluate the achievable communication performance using a more realistic \emph{evaluation} model that employs the \emph{ground-truth} channel parameters and explicitly accounts for practical compensation imperfections. This separation is intentional: it enables a direct assessment of the effectiveness of \ac{pace} by examining how well beamformers optimized with nominal (\ac{pace}-estimated) parameters perform when deployed in a realistic setting.

Following the Doppler-compensation-free assumption adopted in\cite{poor2024tsp,moewin2025jsac}, we do not rely on explicit Doppler precompensation at the transmitter in the evaluation model for simplicity. Meanwhile, we account for imperfect delay compensation, which induces residual phase rotations across subcarriers. In addition, motivated by~\cite{hamza2018twc,asyn2024twc,asyn2025tvt}, we further incorporate the resulting \emph{asynchronous interference} due to link-dependent delay mismatch across users in multi-\ac{leo} cooperative transmission.

Specifically, after delay precompensation, the received signal of $u$-th \ac{ut} on $k$-th subcarrier is modeled as\footnote{Note that we assume the difference between $\tilde{\tau}_{s,m}$ and $\tau_{s,m}$ is smaller than the cyclic prefix (CP) length.}
\begin{align}
&y_u\!\left[k\right]
= \sum_{s=1}^{S} e^{-\jmath 2 \pi k\Delta f\left(\tau_{s,u} - \tilde{\tau}_{s,u}\right)}
\bar{\mathbf{h}}_{s,u}^{\mathsf{T}}[k] \mathbf{w}_{s,u}\, s_u \!\left[k\right] \label{eq:eval_signal_model} \\
&\quad + \sum_{m \neq u}^{U} \sum_{s=1}^{S} e^{-\jmath 2 \pi k\Delta f\left(\tau_{s,u} - \tilde{\tau}_{s,m}\right)}
\bar{\mathbf{h}}_{s,u}^{\mathsf{T}}[k] \mathbf{w}_{s,m}\, s_{m}\!\left[k\right] + z_u \!\left[k\right],\nonumber
\end{align}
where $\bar{\mathbf{h}}_{s,u}[k] = \alpha_{s,u}G(\theta_{s,u}^{\mathrm{el}})\mathbf{a}\!\left(\bm{\theta}_{s,u}\right)$ denotes the (approximately) frequency-flat array response term, while the residual delay mismatch is explicitly captured by the subcarrier-dependent phase rotations in~\eqref{eq:eval_signal_model}. 

Accordingly, the achievable rate of $u$-th UT under the evaluation model is given by \eqref{actual_rate_model} at the bottom of this page, which makes explicit that asynchronous interference persists due to \emph{link-dependent delay mismatch across users}, even after applying delay precompensation.

\section{Numerical Results}\label{numer_result}
\subsection{Simulation Parameters}

We model the Earth as a sphere with a radius of $6400\text{ km}$ and adopt a spherical coordinate system centered at the Earth's core. The \acp{ut} are randomly distributed in a circular service area on the Earth's surface with a radius of $200 \text{ km}$~\cite{moewin2023jsac}. The participating \ac{leo} satellites are positioned within a corresponding circular area in space, centered along the extension of the line connecting the Earth's core to the center of the surface service area, at an orbital altitude of $500\text{ km}$. Each satellite travels along an orbital path intersecting this line at a constant speed $\left|\mathbf{v}_s\right| = 7.6\text{ km/s}$, with velocity vectors $\mathbf{v}_s$ tangent to their orbits and \acp{upa} oriented toward Earth's core.
Due to imperfect time-domain and frequency-domain synchronization, the clock bias (respectively, \ac{cfo}) between each satellite and the \ac{ut} is randomly set to a value between $8 \text{ ns}$ and $12 \text{ ns}$ (respectively, between $\Delta f/120$ and $\Delta f/80$), which remains fixed throughout the simulation. To eliminate satellite handover involvement, we consider one large-timescale position-coherent frame containing $100$ small-timescale channel gain-coherent subframes, where the channel gain changes independently from one subframe to another. Throughout this frame, the satellite constellation topology remains constant for simplicity. The downlink positioning and uplink channel estimation consist of $L_{\text{p}} = 10000$ and $L_{\text{c}} =1000$ pilot symbols, respectively.

The channel gain from each \ac{leo} satellite to each \ac{ut}, expressed by \eqref{chan_gain_formula}, is independently generated in each channel-gain coherent slot. The random phase $\psi_{s,u}$ is uniformly cooperative in $[0,2\pi]$. The components that constitute the path loss $\beta_{s,u}$ are generated according to \eqref{large_loss_component}. The free-space path loss is given by $\beta^{\text{FS}}_{s,u} = 20\log_{10}(\|\mathbf{p}_u-\mathbf{p}_s\|) + 20\log_{10}(f_c) -147.55 \text{ [dB]}$\cite{3gpp.38.811}. The atmospheric absorption $\beta^{\text{AB}}_{s,u}$, which depends on signal frequency and satellite elevation angle, is calculated according to \ac{itu} recommendations \cite{ITU676_12} and configured using an off-the-shelf MATLAB script\cite{pinjun2024jsac}. In the considered \ac{los}-dominant suburban/rural/remote scenarios, the clutter loss $\beta^{\text{CL}}_{s,u}$ and the shadow fading $\beta^{\text{SF}}_{s,u}$ originating from near-surface structures are typically weak\cite{pathloss2021cl}. Therefore, we set $\beta^{\text{CL}}_{s,u} = 0\text{ dB}$ and $\beta^{\text{SF}}_{s,u} = 0\text{ dB}$ for convenience. Additionally, ionospheric scintillation can be ignored for frequencies above $6\text{ GHz}$\cite{3gpp.38.811}. For $\beta^{\text{CL}}_{s,u}$ caused by tropospheric scintillation, which is generally difficult to model analytically, we adopt the example suggested in \cite{pinjun2024jsac} as a reference value. The remaining simulation parameters are listed in Table \ref{simu_para}.

\begin{table}[t]
    \centering
    \caption{Simulation Parameters}
    \label{simu_para}
    \begin{tabular}{@{}ll@{}}
        \toprule
        \textbf{Parameter} & \textbf{Value} \\ 
        \midrule
        Central carrier frequency $f_c$ & $12.7 \text{ GHz}$ (Ku band) \\
        Subcarrier spacing $\Delta f$ & $120 \text{ KHz}$ \\
        Subcarrier number $K$ & 1024 \\
        Power budget at each \ac{leo} satellite & $50 \text{ dBm}$\\
        Power budget at each \ac{ut} & $40 \text{ dBm}$\\
        \ac{psd} $N_0$ & $-173.855 \text{ dBm/Hz}$ \\
        Noise figure $F$ & $10 \text{ dB}$\\
        Number of \ac{leo} satellites $S$ & 4\\
        Number of \acp{ut} $U$ & 8\\
        Antenna number $N = N_{\text{h}} \times N_{\text{v}}$ & $16 \times  16$\\
        Antenna radiation gain $G(\cdot)$ & $\sqrt{\frac{3}{4\pi}}\cos(\theta)$\cite{balanis2005antenna}\\
        \bottomrule
    \end{tabular}
\end{table}

\subsection{Compared Schemes}
\subsubsection{Downlink Positioning}
Since we do not optimize the beamforming matrix $\mathbf{F}_s$, it is generated directly based on two principles, which differ depending on whether position information is utilized:  
\begin{itemize}  
    \item \ac{pab}: Leveraging the position information of \acp{ut} obtained from the previous position-coherent frame, each \ac{leo} satellite constructs $\mathbf{F}_s$ as a matrix whose columns correspond to the conjugates of the steering vectors for the \acp{ut}' estimated directions. Specifically, $\mathbf{F}_s = \bar{\zeta}_s [\mathbf{a}^{*}(\hat{\boldsymbol{\theta}}_{s,1}), \ldots, \mathbf{a}^{*}(\hat{\boldsymbol{\theta}}_{s,U})]$, where $\bar{\zeta}_s$ is a normalization factor that ensures compliance with the power budget.  
    \item \ac{vdb}: In the absence of position information, each \ac{leo} satellite generates a vertical beam directed toward the Earth's core. In this case, the beamforming matrix reduces to a steering vector with an elevation angle of $\pi/2$.  
\end{itemize}

\subsubsection{Uplink Channel Estimation}
Depending on the availability of position information at the \ac{leo} satellites, we consider three types of schemes for evaluating the performance of uplink channel estimation:  
\begin{itemize}  
    \item \ac{pace}: This is the proposed two-timescale hybrid downlink positioning and uplink channel estimation scheme. Its performance is rigorously characterized through \ac{mcrb} analysis.  
    \item Perfect-\ac{pace}: This represents an ideal benchmark where perfect position information of the \acp{ut} is assumed to be available at the \ac{leo} satellites. In this case, the performance of uplink channel estimation is evaluated without model mismatch and analyzed using the \ac{crb}.  
    \item Uplink Channel Estimation (UCE): As previously discussed, in the absence of position information, the channel vector, i.e., $\mathbf{h}_{s,u}[k] = e^{-\jmath 2 \pi k \Delta f \tau_{s,u} } \alpha_{s,u} G(\theta_{s,u}^{\mathrm{el}})\mathbf{a}(\boldsymbol{\theta}_{s,u})$, can be directly estimated in the uplink using a \emph{geometry-agnostic} approach. By defining $\boldsymbol{h}_{s,u} = \alpha_{s,u} G(\theta_{s,u}^{\mathrm{el}})\mathbf{a}(\boldsymbol{\theta}_{s,u})$, we analyze the performance limits of estimating $\boldsymbol{h}_{s,u}$ and $\tau_{s,u}$ simultaneously by deriving the \ac{crb}. This derivation extends the approach in \cite{simoCE2008cl}, which was originally developed for a single-carrier system.  
\end{itemize}  

\subsubsection{Downlink Communication}  
To evaluate downlink communication performance, we use the estimated \ac{csi} from the channel estimation schemes above to perform cooperative beamforming, as outlined in Algorithm~\ref{wmmse_bf_opt} (since the satellites only have access to estimated \ac{csi}). The resulting beamformer is then tested with ground-truth \ac{csi} to obtain the actual sum rate. We also consider two benchmark schemes:
\begin{itemize}  
    \item \ac{pab}: This approach is the same as the one introduced for downlink positioning.  
    \item No-Cooperation (NC): In this method, each \ac{ut} is served only by the single \ac{leo} satellite with the strongest signal strength, thereby avoiding inter-satellite cooperation. The \ac{wmmse} algorithm is applied independently at each satellite to optimize its beamformers based solely on its own \ac{csi} for the associated \acp{ut}, while signals from other satellites are treated as interference.
\end{itemize}  

\subsection{Downlink Positioning}

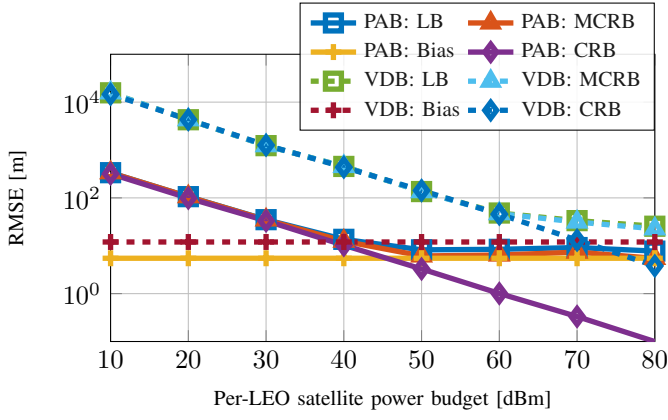
\begin{figure}[t]
\centering 
\centerline{% This file was created by matlab2tikz.
%
%The latest updates can be retrieved from
%  http://www.mathworks.com/matlabcentral/fileexchange/22022-matlab2tikz-matlab2tikz
%where you can also make suggestions and rate matlab2tikz.
%
\definecolor{mycolor1}{rgb}{0.00000,0.44700,0.74100}%
\definecolor{mycolor2}{rgb}{0.85000,0.32500,0.09800}%
\definecolor{mycolor3}{rgb}{0.92900,0.69400,0.12500}%
\definecolor{mycolor4}{rgb}{0.49400,0.18400,0.55600}%
\definecolor{mycolor5}{rgb}{0.46600,0.67400,0.18800}%
\definecolor{mycolor6}{rgb}{0.30100,0.74500,0.93300}%
\definecolor{mycolor7}{rgb}{0.63500,0.07800,0.18400}%
\begin{tikzpicture}

\begin{axis}[%
width=72mm,
height=38mm,
at={(0mm, 0mm)},
scale only axis,
xmin=10,
xmax=80,
xlabel style={font=\color{white!15!black},font=\footnotesize},
xlabel={Per-LEO satellite power budget [dBm]},
ymode=log,
ymin=0.1,
ymax=100000,
yminorticks=true,
ylabel style={font=\color{white!15!black},font=\footnotesize},
ylabel={RMSE [m]},
axis background/.style={fill=white},
xmajorgrids,
ymajorgrids,
% yminorgrids,
legend style={font=\footnotesize, at={(1,1.18)}, anchor=north east, legend cell align=left, align=left, fill = white, fill opacity=0.9, legend columns = 2}
]
legend style={legend cell align=left, align=left, draw=white!15!black}
]
\addplot [color=mycolor1, line width=2.0pt, mark=square, mark size = 3pt, mark options={solid, mycolor1}]
  table[row sep=crcr]{%
10	335.941948892855\\
20	106.094628298932\\
30	35.0773951058647\\
40	13.6222870178463\\
50	8.22276039227371\\
60	8.43080418634218\\
70	9.17601832658302\\
80	7.73388487853722\\
};
\addlegendentry{PAB: LB}

\addplot [color=mycolor2, line width=2.0pt, mark=triangle, mark size = 3pt, mark options={solid, mycolor2}]
  table[row sep=crcr]{%
10	335.897218993336\\
20	105.952908668975\\
30	34.6463901285636\\
40	12.4705815872507\\
50	6.128824519958\\
60	6.40524478502875\\
70	7.35853340397939\\
80	5.45542636667014\\
};
\addlegendentry{PAB: MCRB}

\addplot [color=mycolor3, line width=2.0pt, mark=+, mark size = 3pt, mark options={solid, mycolor3}]
  table[row sep=crcr]{%
10	5.48190646329422\\
20	5.48190646329422\\
30	5.48190646329422\\
40	5.48190646329422\\
50	5.48190646329422\\
60	5.48190646329422\\
70	5.48190646329422\\
80	5.48190646329422\\
};
\addlegendentry{PAB: Bias}

\addplot [color=mycolor4, line width=2.0pt, mark=diamond, mark size = 3pt, mark options={solid, mycolor4}]
  table[row sep=crcr]{%
10	318.430451111055\\
20	98.720440046301\\
30	33.176297797794\\
40	10.2540182601763\\
50	3.28040003226222\\
60	1.00821506712278\\
70	0.332678955653187\\
80	0.0997120793235158\\
};
\addlegendentry{PAB: CRB}

\addplot [color=mycolor5, dashed, line width=2.0pt, mark=square, mark size = 3pt, mark options={solid, mycolor5}]
  table[row sep=crcr]{%
10	15653.8329780253\\
20	4289.39993727687\\
30	1233.79361099758\\
40	453.945304863843\\
50	135.598957533547\\
60	48.2911790109174\\
70	33.504910345628\\
80	25.5846775680663\\
};
\addlegendentry{VDB: LB}

\addplot [color=mycolor6, dashed, line width=2.0pt, mark=triangle, mark size = 3pt, mark options={solid, mycolor6}]
  table[row sep=crcr]{%
10	15653.8283878191\\
20	4289.38318564435\\
30	1233.73537110522\\
40	453.786988779535\\
50	135.068014947487\\
60	46.7795825968369\\
70	31.2869045279073\\
80	22.6023694318037\\
};
\addlegendentry{VDB: MCRB}

\addplot [color=mycolor7, dashed, line width=2.0pt, mark=+, mark size = 3pt, mark options={solid, mycolor7}]
  table[row sep=crcr]{%
10	11.9878531159746\\
20	11.9878531159746\\
30	11.9878531159746\\
40	11.9878531159746\\
50	11.9878531159746\\
60	11.9878531159746\\
70	11.9878531159746\\
80	11.9878531159746\\
};
\addlegendentry{VDB: Bias}

\addplot [color=mycolor1, dashed, line width=2.0pt, mark=diamond, mark size = 3pt, mark options={solid, mycolor1}]
  table[row sep=crcr]{%
10	14608.7630577654\\
20	4328.24024760052\\
30	1263.34059052129\\
40	437.315436002925\\
50	142.827997070129\\
60	45.7547798619192\\
70	13.1315148557393\\
80	3.79798985440482\\
};
\addlegendentry{VDB: CRB}

\end{axis}
\end{tikzpicture}%
}
\caption{Positioning \ac{rmse} \ac{lb} versus per-\ac{leo} satellite power budget.}
% \vspace{-5mm}
\label{Pos_Error_LEO_Power}
\end{figure}

As shown in Fig. \ref{Pos_Error_LEO_Power}, we compare the positioning error bounds and biases under \ac{pab} and \ac{vdb} with different per-satellite power budget. The results reveal that the \ac{lb} of positioning error without model mismatch, represented by the positioning \ac{crb} under perfectly time- and frequency-synchronized satellites, decreases as the transmit power increases. In contrast, when synchronization mismatch is present, the \ac{lb} initially decreases but eventually saturates due to the persistent positioning bias, which remains unaffected by the transmit power. This highlights the detrimental impact of imperfect time- and frequency-synchronization among \ac{leo} satellites on positioning performance \cite{you2024twc,mugen2024tvt}, potentially leading to overly optimistic performance evaluations. Furthermore, \ac{pab} consistently outperforms \ac{vdb} by achieving a lower \ac{lb}, with the performance gain being particularly significant (over an order of magnitude) at lower yet more practical transmit power levels. This advantage stems from \ac{pab}'s ability to leverage position information to allocate more power to the \acp{ut} than \ac{vdb}, underscoring the importance of incorporating prior position knowledge into the design of positioning systems.

\begin{figure}[t]
\centering
\centerline{% This file was created by matlab2tikz.
%
%The latest updates can be retrieved from
%  http://www.mathworks.com/matlabcentral/fileexchange/22022-matlab2tikz-matlab2tikz
%where you can also make suggestions and rate matlab2tikz.
%
\begin{tikzpicture}

\begin{axis}[%
width=72mm,
height=60mm,
at={(0mm, 0mm)},
scale only axis,
xmin=0,
xmax=1e-06,
% xtick={0, 2e-07, 4e-07, 6e-07, 8e-07, 1e-06}, % Add more dense x-axis ticks
tick align=outside,
xlabel style={font=\footnotesize, xshift=0mm,font=\footnotesize},
xlabel={Clock bias [s]},
ymin=0,
ymax=12000,
% ytick={0, 2000, 4000, 6000, 8000, 10000, 12000}, % Add more dense y-axis ticks
ylabel style={font=\footnotesize, yshift=0mm,font=\footnotesize},
ylabel={CFO [Hz]},
zmode=log,
zmin=0,
zmax=1000,
ztick={1, 10, 100, 1000}, % Add more dense z-axis ticks
zlabel style={font=\footnotesize, xshift=0mm,font=\footnotesize},
zlabel={RMSE [m]},
view={-37.5}{30},
axis background/.style={fill=white},
axis x line*=bottom,
axis y line*=left,
axis z line*=left,
xmajorgrids,
ymajorgrids,
zmajorgrids
]

\addplot3[%
surf,
shader=flat, z buffer=sort, colormap={mymap}{[1pt] rgb(0pt)=(0.2422,0.1504,0.6603); rgb(1pt)=(0.25039,0.164995,0.707614); rgb(2pt)=(0.257771,0.181781,0.751138); rgb(3pt)=(0.264729,0.197757,0.795214); rgb(4pt)=(0.270648,0.214676,0.836371); rgb(5pt)=(0.275114,0.234238,0.870986); rgb(6pt)=(0.2783,0.255871,0.899071); rgb(7pt)=(0.280333,0.278233,0.9221); rgb(8pt)=(0.281338,0.300595,0.941376); rgb(9pt)=(0.281014,0.322757,0.957886); rgb(10pt)=(0.279467,0.344671,0.971676); rgb(11pt)=(0.275971,0.366681,0.982905); rgb(12pt)=(0.269914,0.3892,0.9906); rgb(13pt)=(0.260243,0.412329,0.995157); rgb(14pt)=(0.244033,0.435833,0.998833); rgb(15pt)=(0.220643,0.460257,0.997286); rgb(16pt)=(0.196333,0.484719,0.989152); rgb(17pt)=(0.183405,0.507371,0.979795); rgb(18pt)=(0.178643,0.528857,0.968157); rgb(19pt)=(0.176438,0.549905,0.952019); rgb(20pt)=(0.168743,0.570262,0.935871); rgb(21pt)=(0.154,0.5902,0.9218); rgb(22pt)=(0.146029,0.609119,0.907857); rgb(23pt)=(0.138024,0.627629,0.89729); rgb(24pt)=(0.124814,0.645929,0.888343); rgb(25pt)=(0.111252,0.6635,0.876314); rgb(26pt)=(0.0952095,0.679829,0.859781); rgb(27pt)=(0.0688714,0.694771,0.839357); rgb(28pt)=(0.0296667,0.708167,0.816333); rgb(29pt)=(0.00357143,0.720267,0.7917); rgb(30pt)=(0.00665714,0.731214,0.766014); rgb(31pt)=(0.0433286,0.741095,0.73941); rgb(32pt)=(0.0963952,0.75,0.712038); rgb(33pt)=(0.140771,0.7584,0.684157); rgb(34pt)=(0.1717,0.766962,0.655443); rgb(35pt)=(0.193767,0.775767,0.6251); rgb(36pt)=(0.216086,0.7843,0.5923); rgb(37pt)=(0.246957,0.791795,0.556743); rgb(38pt)=(0.290614,0.79729,0.518829); rgb(39pt)=(0.340643,0.8008,0.478857); rgb(40pt)=(0.3909,0.802871,0.435448); rgb(41pt)=(0.445629,0.802419,0.390919); rgb(42pt)=(0.5044,0.7993,0.348); rgb(43pt)=(0.561562,0.794233,0.304481); rgb(44pt)=(0.617395,0.787619,0.261238); rgb(45pt)=(0.671986,0.779271,0.2227); rgb(46pt)=(0.7242,0.769843,0.191029); rgb(47pt)=(0.773833,0.759805,0.16461); rgb(48pt)=(0.820314,0.749814,0.153529); rgb(49pt)=(0.863433,0.7406,0.159633); rgb(50pt)=(0.903543,0.733029,0.177414); rgb(51pt)=(0.939257,0.728786,0.209957); rgb(52pt)=(0.972757,0.729771,0.239443); rgb(53pt)=(0.995648,0.743371,0.237148); rgb(54pt)=(0.996986,0.765857,0.219943); rgb(55pt)=(0.995205,0.789252,0.202762); rgb(56pt)=(0.9892,0.813567,0.188533); rgb(57pt)=(0.978629,0.838629,0.176557); rgb(58pt)=(0.967648,0.8639,0.16429); rgb(59pt)=(0.96101,0.889019,0.153676); rgb(60pt)=(0.959671,0.913457,0.142257); rgb(61pt)=(0.962795,0.937338,0.12651); rgb(62pt)=(0.969114,0.960629,0.106362); rgb(63pt)=(0.9769,0.9839,0.0805)}, mesh/rows=8]
table[row sep=crcr, point meta=\thisrow{c}] {%
x	y	z	c\\
0	0	1.52154021984639	1.52154021984639\\
0	1200	14.3676987640581	14.3676987640581\\
0	1901.87183095334	23.2647635215929	23.2647635215929\\
0	3014.2637178115	37.4027108003759	37.4027108003759\\
0	4777.28604664197	60.762061029736	60.762061029736\\
0	7571.48813376232	90.2947330070803	90.2947330070803\\
0	12000	138.119245507548	138.119245507548\\
1e-09	0	1.58361613528513	1.58361613528513\\
1e-09	1200	14.139992547574	14.139992547574\\
1e-09	1901.87183095334	22.7964284970211	22.7964284970211\\
1e-09	3014.2637178115	36.6748854047315	36.6748854047315\\
1e-09	4777.28604664197	60.8899538851641	60.8899538851641\\
1e-09	7571.48813376232	90.5087005587262	90.5087005587262\\
1e-09	12000	138.435315059426	138.435315059426\\
3.16227766016838e-09	0	1.80187589152599	1.80187589152599\\
3.16227766016838e-09	1200	15.0080989330794	15.0080989330794\\
3.16227766016838e-09	1901.87183095334	23.2951487098253	23.2951487098253\\
3.16227766016838e-09	3014.2637178115	37.2303039802041	37.2303039802041\\
3.16227766016838e-09	4777.28604664197	59.0027274733102	59.0027274733102\\
3.16227766016838e-09	7571.48813376232	90.0492590692442	90.0492590692442\\
3.16227766016838e-09	12000	139.092644158344	139.092644158344\\
1e-08	0	3.53225097829143	3.53225097829143\\
1e-08	1200	18.0396720157069	18.0396720157069\\
1e-08	1901.87183095334	26.1431949976716	26.1431949976716\\
1e-08	3014.2637178115	41.1096141785891	41.1096141785891\\
1e-08	4777.28604664197	63.6983882173665	63.6983882173665\\
1e-08	7571.48813376232	92.7413390127012	92.7413390127012\\
1e-08	12000	140.399377302099	140.399377302099\\
3.16227766016838e-08	0	13.1937813669207	13.1937813669207\\
3.16227766016838e-08	1200	27.9433833732924	27.9433833732924\\
3.16227766016838e-08	1901.87183095334	37.3319348156064	37.3319348156064\\
3.16227766016838e-08	3014.2637178115	52.2852561179088	52.2852561179088\\
3.16227766016838e-08	4777.28604664197	69.8140714811939	69.8140714811939\\
3.16227766016838e-08	7571.48813376232	98.0932757474346	98.0932757474346\\
3.16227766016838e-08	12000	145.942873573388	145.942873573388\\
1e-07	0	51.8427622345979	51.8427622345979\\
1e-07	1200	59.0737303405642	59.0737303405642\\
1e-07	1901.87183095334	64.398254940045	64.398254940045\\
1e-07	3014.2637178115	73.4704601801594	73.4704601801594\\
1e-07	4777.28604664197	88.1805282562391	88.1805282562391\\
1e-07	7571.48813376232	115.442552098673	115.442552098673\\
1e-07	12000	162.320647139885	162.320647139885\\
3.16227766016838e-07	0	119.570364236529	119.570364236529\\
3.16227766016838e-07	1200	127.674971030472	127.674971030472\\
3.16227766016838e-07	1901.87183095334	132.801546761733	132.801546761733\\
3.16227766016838e-07	3014.2637178115	141.483701171478	141.483701171478\\
3.16227766016838e-07	4777.28604664197	155.21362452304	155.21362452304\\
3.16227766016838e-07	7571.48813376232	181.666386999326	181.666386999326\\
3.16227766016838e-07	12000	224.938323337868	224.938323337868\\
1e-06	0	340.035120310095	340.035120310095\\
1e-06	1200	348.656403999536	348.656403999536\\
1e-06	1901.87183095334	353.830843495562	353.830843495562\\
1e-06	3014.2637178115	362.235190316073	362.235190316073\\
1e-06	4777.28604664197	375.90578487542	375.90578487542\\
1e-06	7571.48813376232	398.713958225551	398.713958225551\\
1e-06	12000	436.992779404528	436.992779404528\\
};
\end{axis}
\end{tikzpicture}%
}
\caption{Positioning \ac{rmse} \ac{lb} versus maximum clock bias and \ac{cfo}.}\label{Pos_Error_Bias}
\end{figure}
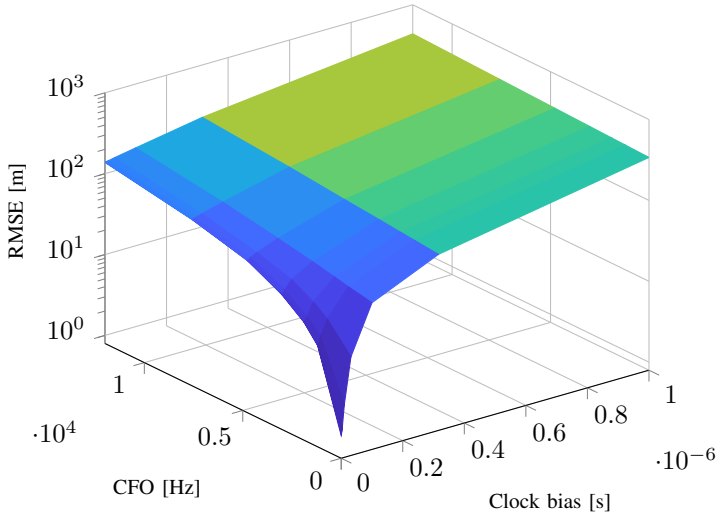

In Fig.~\ref{Pos_Error_Bias}, we analyze the impact of varying clock bias and \ac{cfo} among satellite–\ac{ut} pairs on the \ac{lb} of positioning error (under \ac{pab}), where both impairments are assumed to be uniformly distributed between $0$ and their respective maximum values. The results show that positioning performance degrades significantly as either factor increases. In particular, even a clock bias on the order of tens to hundreds of nanoseconds or a \ac{cfo} of several kHz can lead to positioning errors exceeding $100\text{ m}$. These findings highlight the comparable severity of both impairments and underscore the critical importance of \emph{mitigating mismatches in both the time and frequency domains} to ensure high-accuracy LEO positioning.

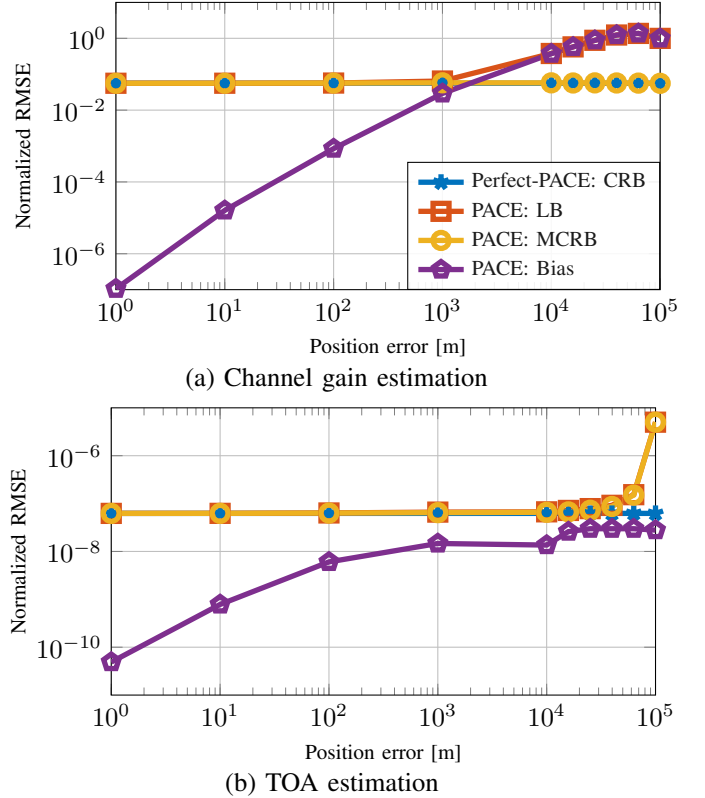
\begin{figure}[htb]
% \vspace{-0.1cm}
\centering
\begin{minipage}[b]{0.98\linewidth}
\vspace{-0.3cm}
  \centering
    % This file was created by matlab2tikz.
%
%The latest updates can be retrieved from
%  http://www.mathworks.com/matlabcentral/fileexchange/22022-matlab2tikz-matlab2tikz
%where you can also make suggestions and rate matlab2tikz.
%
\definecolor{mycolor1}{rgb}{0.00000,0.44700,0.74100}%
\definecolor{mycolor2}{rgb}{0.85000,0.32500,0.09800}%
\definecolor{mycolor3}{rgb}{0.92900,0.69400,0.12500}%
\definecolor{mycolor4}{rgb}{0.49400,0.18400,0.55600}%
\begin{tikzpicture}

\begin{axis}[%
width=72mm,
height=38mm,
at={(0mm,0mm)},
scale only axis,
xmin=1,
xmax=100000,
xmode=log,
xlabel style={font=\color{white!15!black},font=\footnotesize},
xlabel={Position error [m]},
ymode=log,
ymin=1e-07,
ymax=10,
yminorticks=true,
ylabel style={font=\color{white!15!black},font=\footnotesize},
ylabel={Normalized RMSE},
axis background/.style={fill=white},
xmajorgrids,
ymajorgrids,
yminorgrids,
legend style={font=\footnotesize, at={(1,0.445)}, anchor=north east, legend cell align=left, align=left, fill = white, fill opacity=0.9, legend columns = 1}
]
\addplot [color=mycolor1, line width=2.0pt, mark=star, mark size = 3pt]
  table[row sep=crcr]{%
1	0.0560347385198688\\
10	0.0560347385198688\\
100	0.0560347385198688\\
1000	0.0560347385198688\\
10000	0.0560347385198688\\
15848.9319246111	0.0560347385198688\\
25118.8643150958	0.0560347385198688\\
39810.7170553497	0.0560347385198688\\
63095.7344480193	0.0560347385198688\\
100000	0.0560347385198688\\
};
\addlegendentry{Perfect-PACE: CRB}

\addplot [color=mycolor2, line width=2.0pt, mark=square, mark size = 3pt]
  table[row sep=crcr]{%
1	0.0560367311454631\\
10	0.0561322062767975\\
100	0.0567687563295989\\
1000	0.0646584158914307\\
10000	0.377777026113361\\
15848.9319246111	0.577488555185462\\
25118.8643150958	0.880689889172112\\
39810.7170553497	1.23522082517738\\
63095.7344480193	1.36902757272786\\
100000	0.99509646368423\\
};
\addlegendentry{PACE: LB}

\addplot [color=mycolor3, line width=2.0pt, mark=o, mark size = 3pt]
  table[row sep=crcr]{%
1	0.0560367311453631\\
10	0.056132203962359\\
100	0.0567624334977363\\
1000	0.0576814639811737\\
10000	0.0575060145764243\\
15848.9319246111	0.0570145487535998\\
25118.8643150958	0.057379134830938\\
39810.7170553497	0.0570389805095683\\
63095.7344480193	0.0565498091504889\\
100000	0.0557592192510422\\
};
\addlegendentry{PACE: MCRB}

\addplot [color=mycolor4, line width=2.0pt, mark=pentagon, mark size = 3pt]
  table[row sep=crcr]{%
1	1.05825399889547e-07\\
10	1.61192143884666e-05\\
100	0.000847253577414222\\
1000	0.0292157399115233\\
10000	0.373374530125705\\
15848.9319246111	0.574667184203706\\
25118.8643150958	0.878818704725861\\
39810.7170553497	1.2339031735328\\
63095.7344480193	1.36785913528191\\
100000	0.99353302990156\\
};
\addlegendentry{PACE: Bias}

\end{axis}
\end{tikzpicture}%
    \vspace{-1.cm}
  \centerline{(a) Channel gain estimation} \medskip
\end{minipage}
\hfill
\begin{minipage}[b]{0.98\linewidth}
% \vspace{-0.3cm}
  \centering
    % This file was created by matlab2tikz.
%
%The latest updates can be retrieved from
%  http://www.mathworks.com/matlabcentral/fileexchange/22022-matlab2tikz-matlab2tikz
%where you can also make suggestions and rate matlab2tikz.
%
\definecolor{mycolor1}{rgb}{0.00000,0.44700,0.74100}%
\definecolor{mycolor2}{rgb}{0.85000,0.32500,0.09800}%
\definecolor{mycolor3}{rgb}{0.92900,0.69400,0.12500}%
\definecolor{mycolor4}{rgb}{0.49400,0.18400,0.55600}%
\begin{tikzpicture}

\begin{axis}[%
width=72mm,
height=38mm,
at={(0mm,0mm)},
scale only axis,
xmode = log,
xmin=1,
xmax=100000,
xlabel style={font=\color{white!15!black},font=\footnotesize},
xlabel={Position error [m]},
ymode=log,
ymin=1e-11,
ymax=1e-05,
yminorticks=true,
ylabel style={font=\color{white!15!black},font=\footnotesize},
ylabel={Normalized RMSE},
axis background/.style={fill=white},
xmajorgrids,
ymajorgrids,
yminorgrids,
legend style={at={(0.679,0.235)}, anchor=south west, legend cell align=left, align=left, draw=white!15!black}
]
\addplot [color=mycolor1, line width=2.0pt, mark=star, mark size = 3pt]
  table[row sep=crcr]{%
1	6.23788377351158e-08\\
10	6.23788377351158e-08\\
100	6.23788377351158e-08\\
1000	6.23788377351158e-08\\
10000	6.23788377351158e-08\\
15848.9319246111	6.23788377351158e-08\\
25118.8643150958	6.23788377351158e-08\\
39810.7170553497	6.23788377351158e-08\\
63095.7344480193	6.23788377351158e-08\\
100000	6.23788377351158e-08\\
};
% \addlegendentry{CRB}

\addplot [color=mycolor2, line width=2.0pt, mark=square, mark size = 3pt]
  table[row sep=crcr]{%
1	6.2381663488516e-08\\
10	6.25196380244619e-08\\
100	6.36812789443256e-08\\
1000	6.63490379083281e-08\\
10000	6.72380528813979e-08\\
15848.9319246111	7.19149351074594e-08\\
25118.8643150958	7.8510680688065e-08\\
39810.7170553497	9.20135978099782e-08\\
63095.7344480193	1.53410834683695e-07\\
100000	5.01442256005243e-06\\
};
% \addlegendentry{LB}

\addplot [color=mycolor3, line width=2.0pt, mark=o, mark size = 3pt]
  table[row sep=crcr]{%
1	6.23816443193752e-08\\
10	6.25147488515279e-08\\
100	6.33935422915039e-08\\
1000	6.47056633381999e-08\\
10000	6.58499538181037e-08\\
15848.9319246111	6.70609194705923e-08\\
25118.8643150958	7.26170548070958e-08\\
39810.7170553497	8.70499753261138e-08\\
63095.7344480193	1.5048134701441e-07\\
100000	5.01434482744073e-06\\
};
% \addlegendentry{MCRB}

\addplot [color=mycolor4, line width=2.0pt, mark=pentagon, mark size = 3pt]
  table[row sep=crcr]{%
1	4.89040429071543e-11\\
10	7.81866190683286e-10\\
100	6.04682426735886e-09\\
1000	1.46755566614793e-08\\
10000	1.35918849846247e-08\\
15848.9319246111	2.59728891590603e-08\\
25118.8643150958	2.98444355491362e-08\\
39810.7170553497	2.98128156613795e-08\\
63095.7344480193	2.98370306678911e-08\\
100000	2.79206425626353e-08\\
};
% \addlegendentry{Bias}

\end{axis}
\end{tikzpicture}%
    \vspace{-1.cm}
  \centerline{(b) \ac{toa} estimation} \medskip
\end{minipage}
\vspace{-0.4cm}
\caption{Uplink channel estimation \ac{rmse} \ac{lb} versus positioning error: (a) Channel vector estimation; (b) \ac{toa} estimation.}
\label{up_ce_vs_pos_error}
\vspace{-0.2cm}
\end{figure}

\begin{figure}[htb]
% \vspace{-0.1cm}
\centering
\begin{minipage}[b]{0.98\linewidth}
\vspace{-0.3cm}
  \centering
    % This file was created by matlab2tikz.
%
%The latest updates can be retrieved from
%  http://www.mathworks.com/matlabcentral/fileexchange/22022-matlab2tikz-matlab2tikz
%where you can also make suggestions and rate matlab2tikz.
%
\definecolor{mycolor1}{rgb}{0.00000,0.44700,0.74100}%
\definecolor{mycolor2}{rgb}{0.85000,0.32500,0.09800}%
\definecolor{mycolor3}{rgb}{0.92900,0.69400,0.12500}%
\begin{tikzpicture}

\begin{axis}[%
width=72mm,
height=38mm,
at={(0in,0in)},
scale only axis,
xmin=20,
xmax=70,
xlabel style={font=\footnotesize, xshift=0mm},
xlabel={Per-UT power budget [dBm]},
ymode=log,
ymin=0.001,
ymax=10,
yminorticks=true,
ylabel style={font=\footnotesize, yshift=0mm,font=\footnotesize},
ylabel={Normalized RMSE},
axis background/.style={fill=white},
xmajorgrids,
ymajorgrids,
% yminorgrids,
% title={Channel (h) Estimation},
legend style={font=\footnotesize, at={(1,1.16)}, anchor=north east, legend cell align=left, align=left, fill = white, fill opacity=0.9, legend columns = 3}
]
\addplot [color=mycolor1, line width=2.0pt, mark=star, mark size = 3pt]
  table[row sep=crcr]{%
20	0.620097918419685\\
30	0.182139620472802\\
40	0.0488901554548275\\
50	0.0185677304408302\\
60	0.00509647928433054\\
70	0.00183827376120236\\
};
\addlegendentry{PACE}

\addplot [color=mycolor2, line width=2.0pt, mark=square, mark size = 3pt]
  table[row sep=crcr]{%
20	0.567998920888094\\
30	0.181221096101115\\
40	0.0527761251229883\\
50	0.0164123872989707\\
60	0.0055203755861445\\
70	0.00178546198623695\\
};
\addlegendentry{Perfect-PACE}

\addplot [color=mycolor3, line width=2.0pt, mark=o, mark size = 3pt]
  table[row sep=crcr]{%
20	5.6725969398571\\
30	1.80450194561237\\
40	0.567515582139644\\
50	0.179605799113197\\
60	0.0568635680971373\\
70	0.0178455761518213\\
};
\addlegendentry{UCE}

\end{axis}
\end{tikzpicture}%
    \vspace{-1.cm}
  \centerline{(a) Channel vector estimation} \medskip
\end{minipage}
\hfill
\begin{minipage}[b]{0.98\linewidth}
% \vspace{-0.3cm}
  \centering
    % This file was created by matlab2tikz.
%
%The latest updates can be retrieved from
%  http://www.mathworks.com/matlabcentral/fileexchange/22022-matlab2tikz-matlab2tikz
%where you can also make suggestions and rate matlab2tikz.
%
\definecolor{mycolor1}{rgb}{0.00000,0.44700,0.74100}%
\definecolor{mycolor2}{rgb}{0.85000,0.32500,0.09800}%
\definecolor{mycolor3}{rgb}{0.92900,0.69400,0.12500}%
\begin{tikzpicture}

\begin{axis}[%
width=72mm,
height=38mm,
at={(0in,0in)},
scale only axis,
xmin=20,
xmax=70,
xlabel style={font=\footnotesize, xshift=0mm},
xlabel={Per-UT power budget [dBm]},
ymode=log,
ymin=8.37225424102948e-11,
ymax=1e-07,
yminorticks=true,
ylabel style={font=\footnotesize, yshift=0mm,font=\footnotesize},
ylabel={Normalized RMSE},
axis background/.style={fill=white},
xmajorgrids,
ymajorgrids,
% yminorgrids,
% legend style={legend cell align=left, align=left, draw=white!15!black}
]
\addplot [color=mycolor1, line width=2.0pt, mark=star, mark size = 3pt]
  table[row sep=crcr]{%
20	2.64771142496071e-08\\
30	8.37281348820416e-09\\
40	2.64776290109269e-09\\
50	8.3744411611637e-10\\
60	2.65290547022412e-10\\
70	8.53564116607871e-11\\
};
% \addlegendentry{PACE}

\addplot [color=mycolor2, line width=2.0pt, mark=square, mark size = 3pt]
  table[row sep=crcr]{%
20	2.64753925516582e-08\\
30	8.37225424102942e-09\\
40	2.64753925516583e-09\\
50	8.37225424102935e-10\\
60	2.6475392551658e-10\\
70	8.37225424102972e-11\\
};
% \addlegendentry{Perfect-PACE}

\addplot [color=mycolor3, line width=2.0pt, mark=o, mark size = 3pt]
  table[row sep=crcr]{%
20	2.64753925516575e-08\\
30	8.37225424102948e-09\\
40	2.64753925516575e-09\\
50	8.37225424102948e-10\\
60	2.64753925516575e-10\\
70	8.37225424102948e-11\\
};
% \addlegendentry{UCE}

\end{axis}
\end{tikzpicture}%
    \vspace{-1.cm}
  \centerline{(b) \ac{toa} estimation} \medskip
\end{minipage}
\vspace{-0.4cm}
\caption{Uplink channel estimation \ac{rmse} \ac{lb} versus per-\ac{ut} power budget: (a) Channel vector estimation; (b) \ac{toa} estimation.}
\label{up_ce_vs_ut_pow}
% \vspace{-0.2cm}
\end{figure}

\subsection{Uplink Channel Estimation}
In Figs. \ref{up_ce_vs_pos_error}(a) and (b), we evaluate the normalized \ac{rmse}, obtained by dividing the \ac{rmse} by the ground-truth parameters, for both channel gain and \ac{toa} estimation using our proposed \ac{pace}, under varying positioning errors. The results are compared with those of Perfect-\ac{pace}, which assumes no model mismatch due to positioning errors. The results reveal that when positioning errors are small, \ac{pace}'s \ac{lb} closely matches that of Perfect-\ac{pace} for both parameters. However, as positioning errors increase, \ac{pace}'s performance deteriorates due to growing model mismatch. For channel gain estimation, we observe that while the \ac{mcrb}-induced error remains relatively insensitive to positioning errors, the bias component increases proportionally with positioning error, eventually becoming the primary contributor to the overall channel gain error. The \ac{toa} estimation exhibits different behavior: its bias initially increases with positioning error before reaching a plateau, whereas the \ac{mcrb}-induced error continues to grow, ultimately becoming the dominant factor in the total \ac{toa} estimation error. 
Nevertheless, the results indicate that channel estimation degrades noticeably only when the positioning error becomes \emph{impractically large} (e.g., exceeding $1\text{ km}$). This observation highlights the robustness of \ac{pace} to positioning imperfections in \ac{los}-dominant scenarios with relatively high satellite altitudes.

In Figs. \ref{up_ce_vs_ut_pow}(a) and (b), we compare the \ac{lb} for the \ac{rmse} of both channel vector and \ac{toa} estimation using the proposed \ac{pace} against those obtained under UCE, versus \ac{ut}'s power budget. Note that \ac{pace} estimates both the channel gain $\alpha_{s,u}$ and the \ac{toa} $\tau_{s,u}$, whereas UCE estimates the channel vector $\boldsymbol{h}_{s,u} = \alpha_{s,u} G(\theta_{s,u}^{\mathrm{el}})\mathbf{a}(\boldsymbol{\theta}_{s,u})$ (excluding the \ac{toa}-induced phase) in addition to $\tau_{s,u}$. To enable a unified comparison, we transform the \ac{lb} for channel gain estimation \ac{rmse} in \ac{pace} into the corresponding \ac{lb} for channel vector estimation \ac{rmse} (using the form $\hat{\boldsymbol{h}}_{s,u} = \hat{\alpha}_{s,u} G(\hat{\theta}_{s,u}^{\mathrm{el}})\mathbf{a}(\hat{\boldsymbol{\theta}}_{s,u})$) through algebraic manipulation. As shown, the \ac{toa} estimation performance of \ac{pace} closely aligns with that of UCE, given that negligible positioning error is incurred in the positioning stage. This alignment occurs because \ac{toa} estimation is independent of whether the other unknown parameter is the channel gain or the channel vector. This independence can be readily demonstrated by deriving the \ac{crb} for \ac{toa} estimation under Perfect-\ac{pace} and UCE, which we omit here due to space limitations. In contrast, \ac{pace} significantly outperforms UCE in channel vector estimation. This superiority is attributed to \ac{pace}'s utilization of position information, which reduces the complexity of estimating a vector (as required by UCE) to that of estimating a scalar, albeit subject to positioning error. Consequently, \ac{pace} achieves higher channel estimation accuracy under the same pilot length and power budget.

\subsection{Donwlink Communication}
In Fig. \ref{sr_vs_ant_plot}, we compare the downlink communication sum rates achieved by various approaches as a function of the number of antennas deployed on each \ac{leo} satellite. It is observed that Algorithm \ref{wmmse_bf_opt} combined with \ac{pace} significantly outperforms UCE and other baseline methods, achieving sum rates that closely approach those obtained with perfect \ac{csi}. This superior performance is attributed to \ac{pace}'s enhanced channel reconstruction accuracy and the benefits of multi-\ac{leo} satellite cooperation. 
In contrast, the sum rate under NC, whether using \ac{pace} or UCE, remains consistently higher than that achieved by \ac{pab}, even though \ac{pab} leverages inter-satellite cooperation. This highlights the critical role of accurate \ac{csi} in enhancing downlink communication efficiency, beyond what can be achieved using position information alone.
Moreover, as the number of antennas per \ac{leo} satellite increases, the performance advantage of the proposed scheme, i.e., the combination of \ac{pace} and Algorithm \ref{wmmse_bf_opt}, becomes increasingly pronounced, demonstrating its potential for enabling high-throughput \ac{leo} satellite communications with large-scale antenna arrays.

\begin{figure}[t]
\centering 
\centerline{% This file was created by matlab2tikz.
%
%The latest updates can be retrieved from
%  http://www.mathworks.com/matlabcentral/fileexchange/22022-matlab2tikz-matlab2tikz
%where you can also make suggestions and rate matlab2tikz.
%
\definecolor{mycolor1}{rgb}{0.00000,0.44700,0.74100}%
\definecolor{mycolor2}{rgb}{0.85000,0.32500,0.09800}%
\definecolor{mycolor3}{rgb}{0.49400,0.18400,0.55600}%
\definecolor{mycolor4}{rgb}{0.46600,0.67400,0.18800}%
\definecolor{mycolor5}{rgb}{0.30100,0.74500,0.93300}%
\definecolor{mycolor6}{rgb}{0.63500,0.07800,0.18400}%
\begin{tikzpicture}

\begin{axis}[%
width=72mm,
height=38mm,
at={(0mm,0mm)},
scale only axis,
xtick={4, 16, 64, 256}, % Define specific ticks
xmin=4,
xmax=256,
% xmode = log,
% ymode = log,
xlabel style={font=\color{white!15!black},font=\footnotesize},
xlabel={Antenna number at each LEO satellite},
ymin=0,
ymax=21,
ylabel style={font=\color{white!15!black},font=\footnotesize},
ylabel={Sum rate [Mbps]},
axis background/.style={fill=white},
xmajorgrids,
ymajorgrids,
yminorgrids,
legend style={font=\footnotesize, at={(0.86,1.00)}, anchor=north east, legend cell align=left, align=left, fill = white, fill opacity=0.9, legend columns = 2}
]

\addplot [color=mycolor1, line width=2.0pt, mark=triangle, mark size=3.0pt, mark options={solid, mycolor1}]
  table[row sep=crcr]{%
4	0.265613148292845\\
16	1.17023966453443\\
64	4.9275952331816\\
256	19.6612797449126\\
};
\addlegendentry{PACE + Algo. 1}

\addplot [color=mycolor2, line width=2.0pt, mark=square, mark size=3.0pt, mark options={solid, mycolor2}]
  table[row sep=crcr]{%
4	0.198481782509275\\
16	0.891655931498406\\
64	3.6014556765872\\
256	14.4896619017689\\
};
\addlegendentry{UCE + Algo. 1}

\addplot [color=mycolor3, dashed, line width=2.0pt, mark size=3.0pt, mark options={solid, mycolor3}]
  table[row sep=crcr]{%
4	0.310532946063424\\
16	1.24144493848167\\
64	4.95908111614756\\
256	19.733422369892\\
};
\addlegendentry{Perfect CSI + Algo. 1}

\addplot [color=mycolor4, dashed, line width=2.0pt, mark=triangle, mark size=3.0pt, mark options={solid, mycolor4}]
  table[row sep=crcr]{%
4	0.0595045657665146\\
16	0.237952293073666\\
64	0.951579894094719\\
256	3.79913883995801\\
};
\addlegendentry{PAB}

\addplot [color=mycolor5, dashed, line width=2.0pt, mark=+, mark size=3.0pt, mark options={solid, mycolor5}]
  table[row sep=crcr]{%
4	0.106425114501674\\
16	0.426182777493856\\
64	1.70240291594517\\
256	6.79298655638015\\
};
\addlegendentry{PACE + NC}

\addplot [color=mycolor6, dashed, line width=2.0pt, mark=x, mark size=3.0pt, mark options={solid, mycolor6}]
  table[row sep=crcr]{%
4	0.0840690909891586\\
16	0.336946625255357\\
64	1.35064271914237\\
256	5.37268629040149\\
};
\addlegendentry{UCE + NC}

\end{axis}
\end{tikzpicture}%
}
\caption{Sum rate versus per-\ac{leo} satellite antenna number.}
\vspace{-3mm}
\label{sr_vs_ant_plot}
\end{figure}
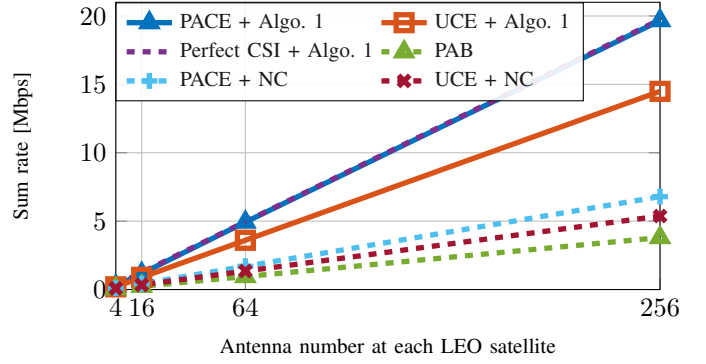

\begin{figure}[t]
\centering 
\centerline{% This file was created by matlab2tikz.
%
%The latest updates can be retrieved from
%  http://www.mathworks.com/matlabcentral/fileexchange/22022-matlab2tikz-matlab2tikz
%where you can also make suggestions and rate matlab2tikz.
%
\definecolor{mycolor1}{rgb}{0.00000,0.44700,0.74100}%
\definecolor{mycolor2}{rgb}{0.85000,0.32500,0.09800}%
\definecolor{mycolor3}{rgb}{0.49400,0.18400,0.55600}%
\definecolor{mycolor4}{rgb}{0.46600,0.67400,0.18800}%
\definecolor{mycolor5}{rgb}{0.30100,0.74500,0.93300}%
\definecolor{mycolor6}{rgb}{0.63500,0.07800,0.18400}%
\begin{tikzpicture}

\begin{axis}[%
width=72mm,
height=38mm,
at={(0mm,0mm)},
scale only axis,
% xtick={4, 16, 64, 256}, % Define specific ticks
xmin=20,
xmax=70,
% xmode = log,
% ymode = log,
xlabel style={font=\color{white!15!black},font=\footnotesize},
xlabel={Per-UT power budget [dBm]},
ymin=0,
ymax=21,
ylabel style={font=\color{white!15!black},font=\footnotesize},
ylabel={Sum rate [Mbps]},
axis background/.style={fill=white},
xmajorgrids,
ymajorgrids,
yminorgrids,
legend style={font=\footnotesize, at={(1.0,0.74)}, anchor=north east, legend cell align=left, align=left, fill = white, fill opacity=0.9, legend columns = 2}
]

\addplot [color=mycolor1, line width=2.0pt, mark=triangle, mark size=3.0pt, mark options={solid, mycolor1}]
  table[row sep=crcr]{%
20	15.5183810452141\\
30	19.5845444652749\\
40	20.4313001192481\\
50	20.4720690326031\\
60	20.5112166314084\\
70	20.5070993845672\\
};
\addlegendentry{PACE + Algo. 1}

\addplot [color=mycolor2, line width=2.0pt, mark=square, mark size=3.0pt, mark options={solid, mycolor2}]
  table[row sep=crcr]{%
20	0.498603036328585\\
30	4.37534907399215\\
40	14.893951928039\\
50	19.8194508166797\\
60	20.4441347494361\\
70	20.5060025095609\\
};
\addlegendentry{UCE + Algo. 1}

\addplot [color=mycolor3, dashed, line width=2.0pt, mark size=3.0pt, mark options={solid, mycolor3}]
  table[row sep=crcr]{%
20	20.5097134917139\\
30	20.5097134917139\\
40	20.5097134917139\\
50	20.5097134917139\\
60	20.5097134917139\\
70	20.5097134917139\\
};
\addlegendentry{Perfect CSI + Algo. 1}

\addplot [color=mycolor4, dashed, line width=2.0pt, mark=triangle, mark size=3.0pt, mark options={solid, mycolor4}]
  table[row sep=crcr]{%
20	3.79920302719041\\
30	3.79890617870318\\
40	3.79906105772446\\
50	3.79891312433175\\
60	3.79915125668721\\
70	3.79866712655196\\
};
\addlegendentry{PAB}

\addplot [color=mycolor5, dashed, line width=2.0pt, mark=+, mark size=3.0pt, mark options={solid, mycolor5}]
  table[row sep=crcr]{%
20	6.78720564159483\\
30	6.78558317307429\\
40	6.78891351306683\\
50	6.79197958105746\\
60	6.7911096545148\\
70	6.79114057756062\\
};
\addlegendentry{PACE + NC}

\addplot [color=mycolor6, dashed, line width=2.0pt, mark=x, mark size=3.0pt, mark options={solid, mycolor6}]
  table[row sep=crcr]{%
20	0.275710989562857\\
30	1.95619246604327\\
40	5.43446099475344\\
50	6.62014563773493\\
60	6.77263857030292\\
70	6.78966298511125\\
};
\addlegendentry{UCE + NC}

\end{axis}
\end{tikzpicture}%
}
\caption{Sum rate versus per-\ac{ut} power budget.}
% \vspace{-5mm}
\label{sr_vs_ut_pow_plot}
\end{figure}
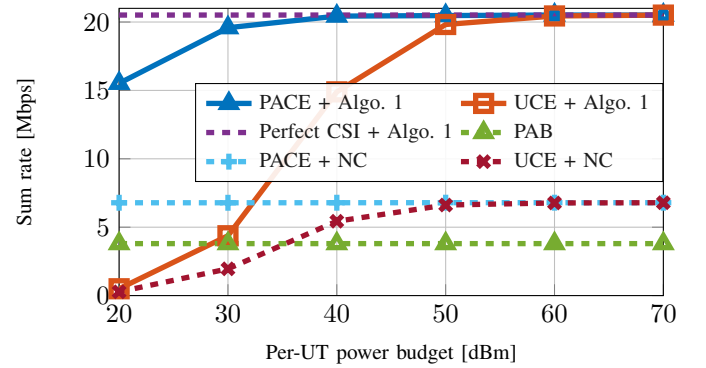
As shown in Fig. \ref{sr_vs_ut_pow_plot}, we compare the sum rates under various schemes versus the per-\ac{ut} power budget. It can be observed that the sum rate under \ac{pab} is independent of the \acp{ut}' transmit power. In contrast, as the \acp{ut}' transmit power increases, the sum rates achieved by Algorithm \ref{wmmse_bf_opt} and NC also increase. This is because higher transmit power improves channel estimation accuracy, which in turn reduces the mismatch between the channel parameters used to optimize beamforming and the ground-truth ones, thereby enhancing the resultant sum rate. Specifically, the sum rates achieved by Algorithm \ref{wmmse_bf_opt} significantly outperform those under NC and \ac{pab}, and gradually approach the performance achieved with perfect \ac{csi}.
This demonstrates the superiority of multi-\ac{leo} satellite cooperative beamforming. Moreover, the use of \ac{pace} results in a substantially higher sum rate compared to UCE, especially when the \acp{ut}' transmit power is low, due to its superior channel estimation accuracy. This highlights the importance of leveraging position information to obtain accurate \ac{csi} when the power budget at the \acp{ut} is limited, an aspect particularly relevant for power-constrained devices such as cell phones.

\section{Conclusion}\label{sec-con}
This paper presents a novel framework for integrating positioning and communication in multi-\ac{leo} satellite systems, leveraging positioning capabilities to enhance channel estimation and thereby facilitate communication. By capitalizing on the distinct timescale variations of position-related parameters and channel gains, we propose a two-timescale frame structure that strategically schedules downlink positioning, uplink channel estimation, and downlink communication. A rigorous \ac{mcrb}-based analysis characterizes the impact of practical imperfections, including inter-satellite clock bias and \ac{cfo}, on positioning accuracy. Additionally, we theoretically demonstrate through \ac{mcrb} analysis how position information improves uplink channel estimation even under inevitable positioning errors. To overcome the limited link budget offered by single-satellite-based downlink communication, we introduce a multi-satellite cooperative beamforming optimization scheme to serve each \ac{ut} simultaneously with multiple satellites within a cluster while reducing complexity. Numerical results validate our theoretical findings. 
 
Beyond the positioning-aided communication emphasized in this work, studying the reverse direction is equally important for future research. For example, inter-\ac{ut} sidelink communication and ranging enable information exchange among proximal \acp{ut} and provide independent timing and distance references. Such mechanisms are promising for mitigating inter-satellite clock biases and thereby improving overall positioning accuracy.

\appendices

\section{MCRB Basics}\label{mcrb_fundamental}
This subsection briefly recaps the fundamentals of \ac{mcrb}.
For a parameter $\bar{\boldsymbol{\theta}}\in \mathbb{R}^{M\times 1}$, the \ac{lb} matrix of the mean squared error of a mismatched estimator is provided by\cite{richmond2017spm}
\addtocounter{equation}{1}
\begin{equation}\label{LBM}
\text{LBM}\big(\tilde{\boldsymbol{\theta}}\big) =  \underbrace{\mathbf{A}^{-1}_{\tilde{\boldsymbol{\theta}}}\mathbf{B}_{\tilde{\boldsymbol{\theta}}}\mathbf{A}^{-1}_{\tilde{\boldsymbol{\theta}}}}_{\text{MCRB}\left(\tilde{\boldsymbol{\theta}}\right)}  + \underbrace{\left(\bar{\boldsymbol{\theta}}-\tilde{\boldsymbol{\theta}}\right)\left(\bar{\boldsymbol{\theta}}-\tilde{\boldsymbol{\theta}}\right)^\top}_{\text{Bias}\left(\tilde{\boldsymbol{\theta}}\right)}. 
\end{equation}
Here, $\tilde{\boldsymbol{\theta}}$ denotes the pseudo-true parameter obtained by
\begin{equation}\label{pseud_para}
\tilde{\boldsymbol{\theta}} = \arg \underset{\boldsymbol{\theta}}{\min\; }   \mathcal{D}\left\{\xi_{\text{T}}\left(\mathbf{y};\bar{\boldsymbol{\theta}}\right)\left|\right|\xi_{\text{M}}\left(\mathbf{y};\boldsymbol{\theta}\right)\right\}, 
\end{equation}
where $\mathcal{D}\{\xi_{\text{T}}\left(\mathbf{y};\boldsymbol{\theta}\right)\left|\right|\xi_{\text{M}}(\mathbf{y};\tilde{\boldsymbol{\theta}})\}$ denotes the Kullback-Leibler (KL) divergence between the true likelihood function $\xi_{\text{T}}\left(\mathbf{y};\boldsymbol{\theta}\right)$ and the mismatched likelihood function $\xi_{\text{M}}(\mathbf{y};\tilde{\boldsymbol{\theta}})$.
In addition, $\mathbf{A}_{\tilde{\boldsymbol{\theta}}}$ and $\mathbf{B}_{\tilde{\boldsymbol{\theta}}}$ represent two generalized \acp{fim}, whose elements in the $i$-th row and the $j$-th column are determined by\cite{richmond2017spm}
\begin{align}\label{GFIM_A}
\left[\mathbf{A}_{\tilde{\boldsymbol{\theta}}}\right]_{i,j}
=&\mathbb{E}_{f_T}\left\{\frac{\partial^2 \ln \xi_M \left(\mathbf{y};\boldsymbol{\theta}\right)}{\partial\left[\boldsymbol{\theta}\right]_i \partial\left[\boldsymbol{\theta}\right]_j}\bigg|_{\boldsymbol{\theta}=\tilde{\boldsymbol{\theta}}}\right\} \notag	\\
=& 2 \Re \left\{\left(\frac{\partial^2 \boldsymbol{\mu}\left(\boldsymbol{\theta}\right)}{\partial\left[\boldsymbol{\theta}\right]_i \partial\left[\boldsymbol{\theta}\right]_j}\right)^{\mathsf{H}} \mathbf{C}_{\text{M}}^{-1}\boldsymbol{\epsilon}\left(\boldsymbol{\theta}\right)  \right.\\
&\;\;\;\;\;\;\;\;\;-\left.\left(\frac{\partial \boldsymbol{\mu}\left(\boldsymbol{\theta}\right)}{\partial\left[\boldsymbol{\theta}\right]_i }\right)^{\mathsf{H}}\mathbf{C}_{\text{M}}^{-1}\left(\frac{\partial \boldsymbol{\mu}\left(\boldsymbol{\theta}\right)}{ \partial\left[\boldsymbol{\theta}\right]_j}\right)\right\}\bigg|_{\boldsymbol{\theta}=\tilde{\boldsymbol{\theta}}}, \notag
\end{align}
and 
\begin{align}\label{GFIM_B}
\left[\mathbf{B}_{\tilde{\boldsymbol{\theta}}}\right]_{i,j}
=&\mathbb{E}_{f_T}\left\{\frac{\partial \ln f_M \left(\mathbf{y};\boldsymbol{\theta}\right)}{\partial\left[\boldsymbol{\theta}\right]_i}\bigg|_{\boldsymbol{\theta}=\tilde{\boldsymbol{\theta}}} \frac{\partial \ln f_M \left(\mathbf{y};\boldsymbol{\theta}\right)}{\partial\left[\boldsymbol{\theta}\right]_j}\bigg|_{\boldsymbol{\theta}=\tilde{\boldsymbol{\theta}}}\right\} \nonumber	\\
=& 4 \Re\left\{\boldsymbol{\epsilon}\left(\boldsymbol{\theta}\right)^{\mathsf{H}}\mathbf{C}_{\text{M}}^{-1}\frac{\partial \boldsymbol{\mu}\left(\boldsymbol{\theta}\right)}{\partial \left[\boldsymbol{\theta}\right]_i}\right\}\Re\left\{\boldsymbol{\epsilon}\left(\boldsymbol{\theta}\right)^{\mathsf{H}} \mathbf{C}_{\text{M}}^{-1} \frac{\partial \boldsymbol{\mu}\left(\boldsymbol{\theta}\right)}{\partial \left[\boldsymbol{\theta}\right]_j}\right\} \nonumber\\
+&2\Re\left\{\left(\frac{\partial \boldsymbol{\mu}\left(\boldsymbol{\theta}\right)}{\partial \left[\boldsymbol{\theta}\right]_i}\right)^{\mathsf{H}} \mathbf{C}_{\text{M}}^{-1} \left(\frac{\partial \boldsymbol{\mu}\left(\boldsymbol{\theta}\right)}{\partial \left[\boldsymbol{\theta}\right]_j}\right)\right\}\bigg|_{\boldsymbol{\theta}=\tilde{\boldsymbol{\theta}}},
\end{align}
respectively. Here, $\mathbf{C}_{\text{M}}$ represents the covariance matrix of $\xi_{\text{M}}(\mathbf{y};\boldsymbol{\theta})$ and $\boldsymbol{\epsilon}(\tilde{\boldsymbol{\theta}}) = \boldsymbol{\kappa}(\bar{\boldsymbol{\theta}}) - \boldsymbol{\mu}(\tilde{\boldsymbol{\theta}})$ with $\boldsymbol{\kappa}(\bar{\boldsymbol{\theta}})$ and $\boldsymbol{\mu}(\tilde{\boldsymbol{\theta}})$ being the noise-free observations under the true and mismatched models, respectively\cite{richmond2017spm,cuneyd2024twc}.

\section{Derivatives of $\tilde{\mathbf{f}}\left(\mathbf{r}_u\right)$ with respect to $\mathbf{r}_u$}\label{deriv_pos}

\emph{1) Non-Zero First-Order Derivative:}
\begin{subequations}
\begin{align}
\frac{\partial \upsilon_{s,u}}{\partial \mathbf{p}_u}   
&=\frac{\mathbf{v}_s}{\lambda\|\mathbf{p}_u-\mathbf{p}_s\|} -\frac{\left(\mathbf{v}_s^{\mathsf{T}}\left(\mathbf{p}_u-\mathbf{p}_s\right)\right)\left(\mathbf{p}_u-\mathbf{p}_s\right)}{\lambda\|\mathbf{p}_u-\mathbf{p}_s\|^3}, \notag \\
\frac{\partial \tau_{s,u}}{\partial \mathbf{p}_u}   
&=\frac{\mathbf{p}_u-\mathbf{p}_s}{c\|\mathbf{p}_u-\mathbf{p}_s\|},\quad \frac{\partial \upsilon_{s,u}}{\partial \delta_u}   =1, \quad 
\frac{\partial \tau_{s,u}}{\partial b_{u}}   
=1.\notag 
\end{align}    
\end{subequations}

\emph{2) Non-Zero Second-Order Derivative:}
\begin{subequations}
\begin{align}
\frac{\partial^2 \upsilon_{s,u}}{\partial \mathbf{p}_u^2} 
&=-\frac{\mathbf{v}_s\left(\mathbf{p}_u-\mathbf{p}_s\right)^{\mathsf{T}}}{\lambda\|\mathbf{p}_u-\mathbf{p}_s\|^3} \notag \\
&-\frac{\mathbf{v}_s\left(\mathbf{p}_u-\mathbf{p}_s\right)^{\mathsf{T}} + \mathbf{v}_s^{\mathsf{T}}\left(\mathbf{p}_u-\mathbf{p}_s\right)\mathbf{I}_3 }{\lambda\|\mathbf{p}_u-\mathbf{p}_s\|^3} \notag \\
&+ \frac{3\left(\mathbf{v}_s^{\mathsf{T}}\left(\mathbf{p}_u-\mathbf{p}_s\right)\right)\left(\mathbf{p}_u-\mathbf{p}_s\right)\left(\mathbf{p}_u-\mathbf{p}_s\right)^{\mathsf{T}}}{\lambda\|\mathbf{p}_u-\mathbf{p}_s\|^5},\notag \\
\frac{\partial^2 \tau_{s,u}}{\partial \mathbf{p}_u^2}   
&=\frac{\mathbf{I}_3}{c\|\mathbf{p}_u-\mathbf{p}_s\|} - \frac{\left(\mathbf{p}_u-\mathbf{p}_s\right)\left(\mathbf{p}_u-\mathbf{p}_s\right)^{\mathsf{T}}}{c\|\mathbf{p}_u-\mathbf{p}_s\|^3}.\notag 
\end{align}    
\end{subequations}

\section{Derivatives of \texorpdfstring{$\tilde{\mathbf{g}}\left(\boldsymbol{\eta}_{s,u}\right)$}{TEXT2} with respect to \texorpdfstring{$\boldsymbol{\eta}_{s,u}$}{TEXT3}}\label{deriv_chan}
Note that the derivatives $\partial\tilde{\mathbf{g}}(\boldsymbol{\eta}_{s,u})/\partial [\boldsymbol{\eta}_{s,u}]_i$ and $\partial^2 \tilde{\mathbf{g}}(\boldsymbol{\eta}_{s,u})/(\partial [\boldsymbol{\eta}_{s,u}]_i \partial [\boldsymbol{\eta}_{s,u}]_j)$ can be constructed by concatenating $\partial\tilde{\mathbf{g}}_{s,u}[\ell,k]/\partial [\boldsymbol{\eta}_{s,u}]_i$ and $\partial^2 \tilde{\mathbf{g}}_{s,u}[\ell,k]/(\partial [\boldsymbol{\eta}_{s,u}]_i \partial [\boldsymbol{\eta}_{s,u}]_j)$, respectively, for $\ell=1,\dots,L_{\text{c}}$ and $k=1,\dots,K$. For simplicity, we provide the latter below.

\emph{1) Non-Zero First-Order Derivative:}
\begin{subequations}
\begin{align}
\frac{\partial \tilde{\mathbf{g}}_{s,u}\left[\ell,k\right]}{\partial \Re\left\{\vartheta{s,u}\right\} }   
&=e^{-\jmath 2\pi k \Delta f \tau_{s,u} } \tilde{\mathbf{d}}_{s,u}\left[\ell,k\right], \notag \\
\frac{\partial \tilde{\mathbf{g}}_{s,u}\left[\ell,k\right]}{\partial \Im\left\{\vartheta_{s,u}\right\} }  
&=\jmath e^{-\jmath 2\pi k \Delta f \tau_{s,u} } \tilde{\mathbf{d}}_{s,u}\left[\ell,k\right], \notag \\
\frac{\partial \tilde{\mathbf{g}}_{s,u}\left[\ell,k\right]}{\partial \tau_{s,u} }  
&=-\vartheta_{s,u} \jmath 2\pi k \Delta f e^{-\jmath 2\pi k \Delta f \tau_{s,u} } \tilde{\mathbf{d}}_{s,u}\left[\ell,k\right]. \notag
\end{align}    
\end{subequations}

\emph{2) Non-Zero Second-Order Derivative:}
\begin{subequations}
\begin{align}
\frac{\partial^2 \tilde{\mathbf{g}}_{s,u}\left[\ell,k\right]}{\partial \Re\left\{\vartheta_{s,u}\right\} \partial \tau_{s,u}}   
&=-\jmath 2\pi k \Delta f e^{-\jmath 2\pi k \Delta f \tau_{s,u} } \tilde{\mathbf{d}}_{s,u}\left[\ell,k\right], \notag \\
\frac{\partial^2 \tilde{\mathbf{g}}_{s,u}\left[\ell,k\right]}{\partial \Im\left\{\vartheta_{s,u}\right\} \partial \tau_{s,u}}  
&=2\pi k \Delta f e^{-\jmath 2\pi k \Delta f \tau_{s,u} } \tilde{\mathbf{d}}_{s,u}\left[\ell,k\right], \notag \\
\frac{\partial^2 \tilde{\mathbf{g}}_{s,u}\left[\ell,k\right]}{\partial \tau_{s,u}^2}  
&=-\vartheta_{s,u}  (2\pi k \Delta f)^2 e^{-\jmath 2\pi k \Delta f \tau_{s,u} } \tilde{\mathbf{d}}_{s,u}\left[\ell,k\right]. \notag
\end{align}    
\end{subequations}

\section{Low-Complexity Method for Solving \eqref{dist_prob_bf_opt}}\label{lc_algo}
\subsection{Extracting Per-Satellite Subproblem}
To reduce complexity of solving \eqref{dist_prob_bf_opt}, we instead adopt a \ac{bcd} strategy, optimizing beamformers for each satellite sequentially while fixing the others. For the $s$-th satellite, the subproblem depends only on $\mathbf{w}_s = [\mathbf{w}_{s,1}^{\mathsf{T}},\ldots,\mathbf{w}_{s,U}^{\mathsf{T}}]^{\mathsf{T}}$ and can be reformulated as
\begin{align}\label{prob_recast_bcd_R1}  
\mathop {\min }\limits_{\mathbf{w}_s} \;\;\; &-2\sum_{u=1}^{U}\Re\left\{\omega_{u} \mu_u\mathbf{h}_{s,u}^{\mathsf{T}} \mathbf{w}_{s,u}\right\} \notag \\
&+ \sum_{u=1}^{U}\sum_{m=1}^{U} \omega_{u} \left|\mu_u \right|^2 \left|\mathbf{h}_{s,u}^{\mathsf{T}} \mathbf{w}_{s,m} + \Omega_{s,u,m}\right|^2 \notag \\  
{\rm{s.t.}}\;\;\;  
&\left\|\mathbf{w}_s\right\|^2 \le \frac{P_s}{K}, 
\end{align}  
where $\Omega_{s,u,m}=\sum_{s^{'} \neq s}^{S}\mathbf{h}_{s^{'},u}^{\mathsf{T}} \mathbf{w}_{s^{'},m}$. 

Let $\mathbf{E}_u = [\mathbf{0}_{N\times N}, \ldots, \underbrace{\mathbf{I}_N}_{\text{The } u\text{-th matrix}},\ldots,\mathbf{0}_{N\times N}]\in \mathbb{R}^{N \times NU}$ be the selection matrix. Through algebraic manipulation, the above can be recast as a convex \ac{qcqp} 
\begin{align}\label{prob_recast_quadratic_R1}  
\mathop {\min }\limits_{\mathbf{w}_s} \;\;\; & \mathbf{w}_s^{\mathsf{H}}\mathbf{Q}_s\mathbf{w}_s  -2\Re\left\{\mathbf{b}_s^{\mathsf{H}} \mathbf{w}_{s}\right\} \notag \\  
{\rm{s.t.}}\;\;\;  
&\left\|\mathbf{w}_s\right\|^2 \le \frac{P_s}{K}, 
\end{align}  
where 
\begin{subequations}
\begin{align}
&\mathbf{R}_s = \sum_{u=1}^{U}\omega_u\left|\mu_u \right|^2\mathbf{h}_{s,u}^*\mathbf{h}_{s,u}^{\mathsf{T}} \succeq \mathbf{0}_{N \times N},\label{Eq_R} \\
&\mathbf{Q}_s = \mathbf{I}_{U} \otimes \mathbf{R}_s, \label{Eq_Q} \\
&\mathbf{b}_s = \sum_{u=1}^{U}\omega_u \mu_u^* \mathbf{E}_u^{\mathsf{T}} \mathbf{h}_{s,u}^* - \sum_{u=1}^{U}\sum_{m=1}^{U}\omega_{u} \left|\mu_u \right|^2 \Omega_{s,u,m} \mathbf{E}_m^{\mathsf{T}} \mathbf{h}_{s,u}^*. \notag
\end{align}
\end{subequations}

\subsection{Exploiting Strong Duality}
It is straightforward to observe that the above problem is a convex \ac{qcqp} involving only $\mathbf{w}_s$. Rather than relying on a convex solver such as CVX, which uses interior-point methods and incurs a complexity of $\mathcal{O}((NU)^3)$, we exploit the problem's strong duality to further reduce computational complexity. Specifically, the Lagrangian of the problem is
\begin{equation}
\mathcal{L}\left(\mathbf{w}_s, \lambda\right) =    \mathbf{w}_s^{\mathsf{H}}\mathbf{Q}_s\mathbf{w}_s  -2\Re\left\{\mathbf{b}_s^{\mathsf{H}} \mathbf{w}_{s}\right\} - \lambda \left(\left\|\mathbf{w}_s\right\|^2 - \frac{P_s}{K}\right), \notag
\end{equation}
where $\lambda \ge 0$ is the Lagrangian multiplier associated with the power constraint. Then, the optimal solutions of \eqref{prob_recast_quadratic_R1} can be derived through examining its Karush–Kuhn–Tucker (KKT) conditions as follows
\begin{subequations}
\begin{align}
\left(\mathbf{Q}_s + \lambda \mathbf{I}_{NU} \right)\mathbf{w}_s &= \mathbf{b}_s, \label{kkt_main_R1}\\
\left\|\mathbf{w}_s\right\| &= \sqrt{\frac{P_s}{K}},\label{kkt_pow_R1}\\
\lambda &\ge 0\label{kkt_lambda_R1}.
\end{align}
\end{subequations}

It follows from \eqref{kkt_main_R1} that 
\begin{equation}\label{optimal_bf_form}
\mathbf{w}_s = \left(\mathbf{Q}_s + \lambda \mathbf{I}_{NU}\right)^{-1} \mathbf{b}_s.    
\end{equation}
Therefore, we must choose $\lambda$ such that $\left\|(\mathbf{Q}_s + \lambda \mathbf{I}_{NU})^{-1} \mathbf{b}_s\right\| = \sqrt{P_s/K}$. Define  
\begin{equation}\label{search_func_R1}
g\left(\lambda\right) = \left\|(\mathbf{Q}_s + \lambda \mathbf{I}_{NU})^{-1} \mathbf{b}_s \right\|^2 - \frac{P_s}{K}.
\end{equation}
Since $||(\mathbf{Q}_s + \lambda_2 \mathbf{I}_{NU})^{-1} \mathbf{b}_s ||= \sqrt{\mathbf{b}_s^{\mathsf{H}}(\mathbf{Q}_s + \lambda_2 \mathbf{I}_{NU})^{-2}\mathbf{b}_s} 
< \sqrt{\mathbf{b}_s^{\mathsf{H}}(\mathbf{Q}_s + \lambda_1 \mathbf{I}_{NU})^{-2}\mathbf{b}_s} 
= ||(\mathbf{Q}_s + \lambda_1 \mathbf{I}_{NU})^{-1} \mathbf{b}_s || $
for $\lambda_2 > \lambda_1 \ge 0$, it follows that $g(\lambda)$ is monotonically decreasing with respect to $\lambda$ whenever $\mathbf{Q}_s \succeq \mathbf{0}_{NU}$, which always holds by examining \eqref{Eq_R} and \eqref{Eq_Q}. Based on this property, a unique $\lambda$ can be efficiently identified using a simple line search method such as the Golden-section search.

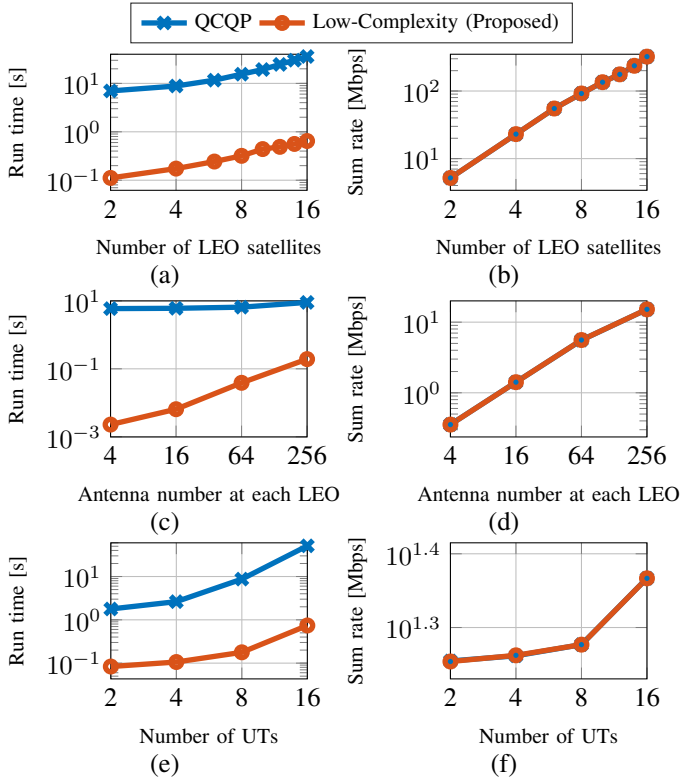
\begin{figure}[h]
\centering
\begin{minipage}[b]{0.49\linewidth}
  \centering
    % This file was created by matlab2tikz.
%
%The latest updates can be retrieved from
%  http://www.mathworks.com/matlabcentral/fileexchange/22022-matlab2tikz-matlab2tikz
%where you can also make suggestions and rate matlab2tikz.
%
\definecolor{mycolor1}{rgb}{0.00000,0.44700,0.74100}%
\definecolor{mycolor2}{rgb}{0.85000,0.32500,0.09800}%
\begin{tikzpicture}

\begin{axis}[%
width=2.6 cm,
height=1.8cm,
at={(0in,0in)},
scale only axis,
xtick={2, 4, 8, 16}, % Define specific ticks
xticklabels={2, 4, 8, 16},
xmode = log,
xmin=2,
xmax=16,
xlabel style={font=\color{white!15!black}, font=\footnotesize},
xlabel={Number of LEO satellites},
ymin=0,
ymax=40,
ymode = log,
ylabel style={font=\color{white!15!black}, font=\footnotesize},
ylabel={Run time [s]},
axis background/.style={fill=white},
xmajorgrids,
ymajorgrids,
% axis x line*=bottom,
% axis y line*=left,
legend style={at={(0.100609,1.08)}, anchor=south west, legend cell align=left, align=left, draw=white!15!black, font=\footnotesize, legend columns = 2}
]
\addplot [color=mycolor1, line width=2.0pt, mark=x, mark size=3.0pt, mark options={solid, mycolor1}]
  table[row sep=crcr]{%
2	6.98604141\\
4	8.82878591811111\\
6	11.6606307764286\\
8	15.4865300008\\
10	19.3431170743333\\
12	24.6472246795\\
14	30.144562307\\
16	35.9094241095\\
};
\addlegendentry{QCQP}

\addplot [color=mycolor2, line width=2.0pt, mark=o, mark options={solid, mycolor2}]
  table[row sep=crcr]{%
2	0.111454407\\
4	0.1738025819\\
6	0.243015318285714\\
8	0.3188733728\\
10	0.439787344\\
12	0.4889468775\\
14	0.561590932\\
16	0.6451663135\\
};
\addlegendentry{Low-Complexity (Proposed)}

\end{axis}
\end{tikzpicture}%
    \vspace{-1 cm}
  \centerline{(a)} \medskip
\end{minipage}
\vspace{-0.3 cm}
\begin{minipage}[b]{0.49\linewidth}
  \centering
    % This file was created by matlab2tikz.
%
%The latest updates can be retrieved from
%  http://www.mathworks.com/matlabcentral/fileexchange/22022-matlab2tikz-matlab2tikz
%where you can also make suggestions and rate matlab2tikz.
%
\definecolor{mycolor1}{rgb}{0.00000,0.44700,0.74100}%
\definecolor{mycolor2}{rgb}{0.85000,0.32500,0.09800}%
\begin{tikzpicture}

\begin{axis}[%
width=2.6 cm,
height=1.8cm,
at={(0in,0in)},
scale only axis,
xtick={2, 4, 8, 16}, % Define specific ticks
xticklabels={2, 4, 8, 16},
xmode = log,
xmin=2,
xmax=16,
xlabel style={font=\color{white!15!black}, font=\footnotesize},
xlabel={Number of LEO satellites},
ymin=0,
ymax=350,
ymode = log,
ylabel style={font=\color{white!15!black}, font=\footnotesize},
ylabel={Sum rate [Mbps]},
axis background/.style={fill=white},
xmajorgrids,
ymajorgrids,
% axis x line*=bottom,
% axis y line*=left,
legend style={at={(0.4,0.78)}, anchor=south west, legend cell align=left, align=left, draw=white!15!black}
]
\addplot [color=mycolor1, line width=2.0pt, mark=x, mark size=3.0pt, mark options={solid, mycolor1}]
  table[row sep=crcr]{%
2	5.17669209353258\\
4	22.9902287214308\\
6	55.073146514619\\
8	91.8524328908872\\
10	135.019911025857\\
12	176.370493158443\\
14	236.120338481777\\
16	322.500351724244\\
};
% \addlegendentry{WMMSE: Actual}

\addplot [color=mycolor2, line width=2.0pt, mark=o, mark options={solid, mycolor2}]
  table[row sep=crcr]{%
2	5.17743935030422\\
4	23.0295576295069\\
6	55.0764791637213\\
8	91.888829035921\\
10	135.031121272905\\
12	176.394168086973\\
14	236.138595829087\\
16	322.556483529699\\
};
% \addlegendentry{WMMSE-LC: Actual}

\end{axis}
\end{tikzpicture}%
    \vspace{-1 cm}
  \centerline{(b)} \medskip
\end{minipage}
\vspace{-0.3 cm}
\begin{minipage}[b]{0.49\linewidth}
  \centering
    % This file was created by matlab2tikz.
%
%The latest updates can be retrieved from
%  http://www.mathworks.com/matlabcentral/fileexchange/22022-matlab2tikz-matlab2tikz
%where you can also make suggestions and rate matlab2tikz.
%
\definecolor{mycolor1}{rgb}{0.00000,0.44700,0.74100}%
\definecolor{mycolor2}{rgb}{0.85000,0.32500,0.09800}%
\begin{tikzpicture}

\begin{axis}[%
width=2.6 cm,
height=1.8cm,
at={(0in,0in)},
scale only axis,
xtick={4, 16, 64, 256}, % Define specific ticks
xticklabels={4, 16, 64, 256},
xmode = log,
xmin=4,
xmax=256,
xlabel style={font=\color{white!15!black}, font=\footnotesize},
xlabel={Antenna number at each LEO},
ymin=0,
ymax=10,
ymode = log,
ylabel style={font=\color{white!15!black}, 
font=\footnotesize},
ylabel={Run time [s]},
axis background/.style={fill=white},
xmajorgrids,
ymajorgrids,
% axis x line*=bottom,
% axis y line*=left,
legend style={legend cell align=left, align=left, draw=white!15!black}
]
\addplot [color=mycolor1, line width=2.0pt, mark=x, mark size=3.0pt, mark options={solid, mycolor1}]
  table[row sep=crcr]{%
4	5.9266934116\\
16	6.05176282958333\\
64	6.530727921125\\
256	9.022352481\\
};
% \addlegendentry{WMMSE}

\addplot [color=mycolor2, line width=2.0pt, mark=o, mark options={solid, mycolor2}]
  table[row sep=crcr]{%
4	0.00228210377777778\\
16	0.00646931458333333\\
64	0.0387501799\\
256	0.193091282\\
};
% \addlegendentry{WMMSE-LC}

\end{axis}
\end{tikzpicture}%
    \vspace{-1 cm}
  \centerline{(c)} \medskip
\end{minipage}
\begin{minipage}[b]{0.49\linewidth}
  \centering
    % This file was created by matlab2tikz.
%
%The latest updates can be retrieved from
%  http://www.mathworks.com/matlabcentral/fileexchange/22022-matlab2tikz-matlab2tikz
%where you can also make suggestions and rate matlab2tikz.
%
\definecolor{mycolor1}{rgb}{0.00000,0.44700,0.74100}%
\definecolor{mycolor2}{rgb}{0.85000,0.32500,0.09800}%
\begin{tikzpicture}

\begin{axis}[%
width=2.6 cm,
height=1.8cm,
at={(0in,0in)},
scale only axis,
xtick={4, 16, 64, 256}, % Define specific ticks
xticklabels={4, 16, 64, 256},
xmode = log,
xmin=4,
xmax=256,
xlabel style={font=\color{white!15!black}, font=\footnotesize},
xlabel={Antenna number at each LEO},
ymin=0,
ymax=20,
ymode = log,
ylabel style={font=\color{white!15!black}, font=\footnotesize},
ylabel={Sum rate [Mbps]},
axis background/.style={fill=white},
xmajorgrids,
ymajorgrids,
% axis x line*=bottom,
% axis y line*=left,
legend style={at={(0.688,0.548)}, anchor=south west, legend cell align=left, align=left, draw=white!15!black}
]
\addplot [color=mycolor1, line width=2.0pt, mark=x, mark size=3.0pt, mark options={solid, mycolor1}]
  table[row sep=crcr]{%
4	0.355342065165409\\
16	1.41329067257316\\
64	5.56533696071263\\
256	15.1877501232294\\
};
% \addlegendentry{WMMSE}

\addplot [color=mycolor2, line width=2.0pt, mark=o, mark options={solid, mycolor2}]
  table[row sep=crcr]{%
4	0.354731767342259\\
16	1.41327803552039\\
64	5.58234768532634\\
256	15.1898532948168\\
};
% \addlegendentry{WMMSE-LC}

\end{axis}
\end{tikzpicture}%
    \vspace{-1 cm}
  \centerline{(d)} \medskip
\end{minipage}
\begin{minipage}[b]{0.49\linewidth}
  \centering
    % This file was created by matlab2tikz.
%
%The latest updates can be retrieved from
%  http://www.mathworks.com/matlabcentral/fileexchange/22022-matlab2tikz-matlab2tikz
%where you can also make suggestions and rate matlab2tikz.
%
\definecolor{mycolor1}{rgb}{0.00000,0.44700,0.74100}%
\definecolor{mycolor2}{rgb}{0.85000,0.32500,0.09800}%
\begin{tikzpicture}

\begin{axis}[%
width=2.6 cm,
height=1.8cm,
at={(0in,0in)},
scale only axis,
xtick={2, 4, 8, 16}, % Define specific ticks
xticklabels={2, 4, 8, 16},
xmode = log,
xmin=2,
xmax=16,
xlabel style={font=\color{white!15!black}, font=\footnotesize},
xlabel={Number of UTs},
ymin=0,
ymax=60,
ymode = log,
ylabel style={font=\color{white!15!black}, font=\footnotesize},
ylabel={Run time [s]},
axis background/.style={fill=white},
xmajorgrids,
ymajorgrids,
% axis x line*=bottom,
% axis y line*=left,
legend style={legend cell align=left, align=left, draw=white!15!black}
]
\addplot [color=mycolor1, line width=2.0pt, mark=x, mark size=3.0pt, mark options={solid, mycolor1}]
  table[row sep=crcr]{%
2	1.78124816666667\\
4	2.63157842983333\\
8	8.7158827585\\
16	51.099675481\\
};
% \addlegendentry{WMMSE}

\addplot [color=mycolor2, line width=2.0pt, mark=o, mark options={solid, mycolor2}]
  table[row sep=crcr]{%
2	0.0834823145\\
4	0.106022650142857\\
8	0.17768415725\\
16	0.740673771\\
};
% \addlegendentry{WMMSE-LC}

\end{axis}
\end{tikzpicture}%
    \vspace{-1 cm}
  \centerline{(e)} \medskip
\end{minipage}
\begin{minipage}[b]{0.49\linewidth}
  \centering
    % This file was created by matlab2tikz.
%
%The latest updates can be retrieved from
%  http://www.mathworks.com/matlabcentral/fileexchange/22022-matlab2tikz-matlab2tikz
%where you can also make suggestions and rate matlab2tikz.
%
\definecolor{mycolor1}{rgb}{0.00000,0.44700,0.74100}%
\definecolor{mycolor2}{rgb}{0.85000,0.32500,0.09800}%
\begin{tikzpicture}

\begin{axis}[%
width=2.6 cm,
height=1.8cm,
at={(0in,0in)},
scale only axis,
xtick={2, 4, 8, 16}, % Define specific ticks
xticklabels={2, 4, 8, 16},
xmode = log,
xmin=2,
xmax=16,
xlabel style={font=\color{white!15!black}, font=\footnotesize},
xlabel={Number of UTs},
ymin=17,
ymax=26,
ymode = log,
ylabel style={font=\color{white!15!black}, font=\footnotesize},
ylabel={Sum rate [Mbps]},
axis background/.style={fill=white},
xmajorgrids,
ymajorgrids,
% axis x line*=bottom,
% axis y line*=left,
legend style={legend cell align=left, align=left, draw=white!15!black}
]
\addplot [color=mycolor1, line width=2.0pt, mark=x, mark size=3.0pt, mark options={solid, mycolor1}]
  table[row sep=crcr]{%
2	17.9789127076081\\
4	18.2708335908704\\
8	18.9194749489495\\
16	23.2749738170899\\
};
% \addlegendentry{WMMSE}

\addplot [color=mycolor2, line width=2.0pt, mark=o, mark options={solid, mycolor2}]
  table[row sep=crcr]{%
2	17.9482469050658\\
4	18.296805456728\\
8	18.9188448923556\\
16	23.269910476239\\
};
% \addlegendentry{WMMSE-LC}

\end{axis}
\end{tikzpicture}%
    \vspace{-1 cm}
  \centerline{(f)} \medskip
\end{minipage}
\vspace{-0.8 cm}
\caption{{Run time and sum rate comparison between the QCQP-based and proposed low-complexity algorithms: (a) Run time versus LEO satellite number; (b) Sum rate versus LEO satellite number; (c) Run time versus antenna number; (d) Sum rate versus antenna number; (e) Run time versus UT number; (f) Sum rate versus UT number.}}
\label{complexity_vs_leo_num}
\vspace{-0.2cm}
\end{figure}

\subsection{Reducing Matrix Inversion Complexity}
We note that \eqref{search_func_R1} involves a $NU$-dimensional matrix inversion, which still incurs a computational complexity of $\mathcal{O}((NU)^3)$ per iteration of the line search. To reduce this complexity, we exploit the structure $\mathbf{Q}_s = \mathbf{I}_{U} \otimes \mathbf{R}_s$. Let $\mathbf{b}_s = [\mathbf{b}_{s,1}^{\mathsf{T}},\ldots,\mathbf{b}_{s,U}^{\mathsf{T}}]^{\mathsf{T}}$, where $\mathbf{b}_{s,u} \in \mathbb{C}^{N}$. We then have
\begin{align}
g\left(\lambda\right) &= \left\|\left(\mathbf{I}_{U} \otimes \mathbf{R}_s + \lambda \mathbf{I}_{NU}\right)^{-1} \mathbf{b}_s \right\|^2 - \frac{P_s}{K} \notag \\
&= \left\|\left(\mathbf{I}_{U} \otimes \left(\mathbf{R}_s + \lambda \mathbf{I}_{N}\right)\right)^{-1} \mathbf{b}_s \right\|^2 - \frac{P_s}{K} \notag \\
& = \left\|\left(\mathbf{I}_{U} \otimes \left(\mathbf{R}_s + \lambda \mathbf{I}_{N}\right)^{-1}\right) \mathbf{b}_s \right\|^2 - \frac{P_s}{K} \notag \\
&= \left\|\left(\mathbf{I}_{U} \otimes \left(\mathbf{U}\left(\mathbf{\Lambda} + \lambda \mathbf{I}_{N}\right)^{-1}\mathbf{U}^{\mathsf{H}}\right)\right) \mathbf{b}_s \right\|^2 - \frac{P_s}{K} \notag \\
& = \sum_{u=1}^{U}\left\|\mathbf{U}\left(\mathbf{\Lambda} + \lambda \mathbf{I}_{N}\right)^{-1}\mathbf{U}^{\mathsf{H}}\mathbf{b}_{s,u}\right\|^2 - \frac{P_s}{K} \notag \\
&= \sum_{u=1}^{U}\sum_{n=1}^{N} \frac{\left|\varpi_{s,u,n}\right|^2}{\left(\lambda_n + \lambda\right)^2}  - \frac{P_s}{K}, \label{final_Search_form}
\end{align}
where $\boldsymbol{\varpi}_{s,u} = [\varpi_{s,u,1},\ldots,\varpi_{s,u,N}]^\mathsf{T} = \mathbf{U}^{\mathsf{H}}\mathbf{b}_{s,u}$. Here, $\mathbf{R}_s$ is decomposed as $\mathbf{R}_s = \mathbf{U} \mathbf{\Lambda} \mathbf{U}^{\mathsf{H}}$ via eigenvalue decomposition, where $\mathbf{U} \in \mathbb{C}^{N \times N}$ is a unitary matrix and $\mathbf{\Lambda} = \text{diag}[\lambda_1, \ldots, \lambda_N]$ is the diagonal eigenvalue matrix. Then, \eqref{optimal_bf_form} is reformulated as
\begin{equation}\label{optimal_bf_form_lc}
\mathbf{w}_s = \left(\mathbf{I}_{U} \otimes \left(\mathbf{U}\left(\mathbf{\Lambda} + \lambda \mathbf{I}_{N}\right)^{-1}\mathbf{U}^{\mathsf{H}}\right)\right) \mathbf{b}_s,
\end{equation}
which eliminates the need for an $NU$-dimensional matrix inversion and thereby reduces the computational complexity of updating beamformers.

\subsection{Analyzing Complexity}
The above reformulation shows that eigenvalue decomposition of the matrix $\mathbf{R}_s$ needs to be performed only once with complexity $\mathcal{O}(N^3)$, rather than repeatedly inverting an $NU$-dimensional matrix for each $\lambda$. The Golden-section search method, which has a linear convergence rate, finds an $\epsilon$-solution within $\mathcal{O}(\log(1/\epsilon))$ iterations. Since each iteration only requires evaluating a one-dimensional function, the associated complexity can generally be neglected. Consequently, the eigenvalue decomposition step dominates the overall complexity of solving \eqref{prob_recast_bcd_R1}. As \eqref{prob_recast_bcd_R1} is solved sequentially for $s=1,\ldots,S$, the total complexity of solving \eqref{dist_prob_bf_opt} is reduced to $\mathcal{O}(SN^3)$, compared with the original $\mathcal{O}((SNU)^3)$.

To further demonstrate the superiority of the proposed algorithm, Fig.~\ref{complexity_vs_leo_num} compares the run time and sum rate of the \ac{qcqp}-based and proposed low-complexity algorithms under varying numbers of \ac{leo} satellites, antennas per \ac{leo}, and \acp{ut}. The results show that the proposed algorithm, which decomposes the beamformer update subproblem via \ac{bcd} and exploits duality and line search, achieves orders-of-magnitude complexity reduction while maintaining identical sum-rate performance compared with the original \ac{qcqp}-based update. This is notable since the \ac{qcqp} formulation yields the optimal subproblem solution, whereas the proposed \ac{bcd}-based structure converges only to a stationary point. These results validate the efficiency of the proposed low-complexity design.

\bibliographystyle{IEEEtran}
\bibliography{IEEEabrv,mybib}

\end{document}